\documentclass[graybox, secnum]{svmult}

\usepackage{mathptmx}      
\usepackage{helvet}            
\usepackage{courier}          
\usepackage{type1cm}        
\usepackage{makeidx}         
\usepackage{graphicx}         
\usepackage{multicol}          
\usepackage[bottom]{footmisc}
\usepackage{hyperref}         
\usepackage{soul}                

\usepackage{amsmath}
\usepackage{amssymb}
\usepackage{amsfonts}
\usepackage{color}
\usepackage{indentfirst}
\usepackage{latexsym}

\newcommand{\dd}{\mathrm d}

\newcommand{\lp}{\ensuremath{\left(}}
\newcommand{\rp}{\ensuremath{\right)}}
\newcommand{\lc}{\ensuremath{\left[}}
\newcommand{\rc}{\ensuremath{\right]}}

\newcommand{\MP}{M_\mathrm{_P}}

\def\IC{\relax\hbox{$\inbar\kern-.3em{\rm C}$}}

\renewcommand{\vec}[1]{{\bf #1}}

\def\beq{\begin{eqnarray}}
\def\eeq{\end{eqnarray}}
\def\nn{\nonumber}                
\def\n{\label}                          

\def\ln{\,\mbox{ln}\,}
\def\Ln{\,\mbox{Ln}\,}
\def\Det{\,\mbox{Det}\,}

\def\tr{\,\mbox{tr}\,}
\def\diag{\,\mbox{diag}\,}
\def\Tr{\,\mbox{Tr}\,}

\def\sign{{\rm sign}}
\def\al{\alpha}
\def\be{\beta}

\def\ga{\gamma}
\def\de{\delta}
\def\vp{\varepsilon}
\def\ep{\epsilon}

\def\ka{\kappa}
\def\la{\lambda}
\def\na{\nabla}
\def\pa{\partial}

\def\si{\sigma}
\def\om{\omega}

\def\ta{\tau}
\def\th{\theta}
\def\te{\vartheta}

\def\Ga{\Gamma}
\def\De{\Delta}
\def\La{\Lambda}
\def\Si{\Sigma}
\def\Om{\Omega}

\DeclareMathOperator{\cx}{\square}

\hypersetup{colorlinks=true,urlcolor=blue}
\usepackage[square,numbers]{natbib}
\makeindex             

\begin{document}
\title*{The background information about perturbative quantum gravity }
\author{Ilya L. Shapiro \thanks{corresponding author}}
\institute{
Departamento de F\'{\i}sica,
Universidade Federal de Juiz de Fora, MG, 36036-900, Brazil
\\
E-mail address: \ ilyashapiro2003@ufjf.br
\\
On leave from Tomsk State Pedagogical University, Russia}
%
%
\maketitle

\abstract{
The purpose of this Chapter is to give a general introduction and
status review on the perturbative approach to quantum gravity (QG).
This text is a modified version of the corresponding chapters of
Part II of the recent textbook on quantum field theory (QFT) and
QG, co-authored with I.L.~Buchbinder and published in Oxford
University Press.
We discuss the choice of the starting action in the QG models,
degrees of freedom and propagator of metric perturbations, power
counting and renormalizability of these models, the problems
related to higher derivative theories and ghosts, such as quantum
unitarity and the stability of classical solutions in general relativity;
and the perspective to overcome these problems.  The gauge fixing
and parametrization dependencies are discussed in detail using the
corresponding general QFT theorems developed in gauge theories.
On top of that, we present a basic example of deriving the one-loop
divergences and discuss an important example of the renormalization
group in QG.
The gauge invariant renormalizability of QG is considered in another
Chapter of the Handbook, written together with P.M. Lavrov.
}

\section*{Keywords}
\noindent
Quantum gravity; gravitational action; gauge fixing; propagator;
power counting; renormalizability; higher derivatives; ghosts; stability.


\section{Introduction}
\label{sec.Intro}

Perturbative quantum gravity (QG) aims to construct quantum
gravitational theory in a manner close to how QFT describes other
fundamental forces, such as electroweak and high-energy strong
interactions. One of the main purposes of
perturbative QG is to get the quantum corrections to the classical
(tree-level) action of gravity and the corresponding effective
equations of motion for the metric.

The importance of loop corrections to the gravitational equations,
coming from QG or from the quantum effects of matter fields
(semiclassical approach), is partially owing to the fact that, at
some point, we hope to be able to compare these corrections with
experimental or observational data. Another reason to study QG
is that the consistency of quantum theory may be useful to establish
the restrictions on the modifications of general relativity (GR).
It is assumed we may be able to select those gravity theories that
are consistent at the quantum level and discard other models.
Independent of all these reasons, in the last six decades the
perturbative QG became an important part of the general QFT
scenario and, especially, of the gauge fields theories.

When initiating the QG, we meet two main choices representing
kinds of the points of bifurcation, where one can choose the
direction of how to construct the theory.
The first question is to decide what should be the object of
quantization. In the traditional perturbative quantum gravity, we
choose to quantize the spacetime metric, and apply to it the rules
of quantization that are common to all gauge theories, such as the
Yang-Mills. It is worth mentioning that one can choose other
quantum variables to describe gravity, e.g., the tetrad or consider
the connection independent of the metric (first order formalism).
Owing to the size and contents limitations, we will not discuss
these possibilities in what follows.

On the other hand, it is
possible to extend the set of the quantum
fields, that is include other fields along with the metric, or even
replacing the metric. One of the most interesting approaches to
make such extension is to provide an extended symmetry, as it is
done in supergravity. We can refer the reader to the corresponding
Section of the present Handbook for the reviews of supergravity,
but will not discuss it here. Another possibility is to add quantum
matter fields. Let us note that there is a reduced, semiclassical
approach, when only the matter fields are assumed quantum and
metric is regarded as a classical background. This kind of theories
share many technical and conceptual features with perturbative QG,
so the reader may find useful to study the semiclassical approach
first. There are many useful monographs on this subject, let us
mention just a few of them,
\cite{DeWitt65,birdav,book,MuWi,ParToms,Toms} and also recent
textbook \cite{OUP}, where the reader can also find an introduction
to QG\footnote{For the interested reader, there are good books
and reviews of semiclassical and quantum gravity theories, e.g.,
\cite{Alvarez-88,Fulling,Kiefer,Percacci}. One can find many
other useful sources starting form these references.}.
In what follows, we shall address the same subjects which
are discussed in sections 18-21 of this textbook. Most of the present
Chapter can be seen as a modified and more review-style extract
from this book.

The second important choice appears after we agree to quantize the
metric. One has to choose the classical theory which should serve as
a basis for the perturbative quantum gravity. Obviously, such initial
model may be the Einstein's GR, however there
are strong reasons to try also other models of gravity. The choice of
a model defines many important features of the theory, including its
particle contents, renormalizability, unitarity, stability of classical
solutions, high energy (UV) and low-energy (IR) behaviors, as we
shall discuss below.

In what follows, we review the main elements of the perturbative
QG, referring to the recent textbook \cite{OUP} for most of the
technical details. First of all, we consider the choice of the
action and the corresponding gauge-fixing conditions, mostly,
in the framework of the background field method
\cite{DeWitt65,Abbott82} (the reader can find more references
starting from these two). After that, we review the structure of
propagator of the metric perturbations in various models and the
corresponding analysis of the degrees of freedom in QG.

The proof of the gauge-invariant renormalizability in QG is left
for the  second chapter of this Section \cite{LavSha-HQG}. This
analysis possesses higher level of complexity compared to the
mentioned textbook, but, as a result, the main statements are
obtained in a more general and more concrete way. The main two
outputs of this consideration, based on the BRST symmetry and
the Batalin-Vilkovisky technique, are the following two statements:

\textit{i)} The counterterms in the generally covariant theory of QG
may have the same symmetry, i.e., the diffeomorphism invariance,
as the initial classical action. This symmetry holds at the quantum
level if we use a regularization preserving this symmetry, such
that the general covariance is not violated by quantum corrections.
For instance, this is the case for the perturbative QG in even
dimension $n=2m+2$ (where $m=1,2,,\dots$), including $n=4$,
the last will be used by default in the rest of this Chapter.

\textit{ii)} The dependence on the choice of the gauge fixing
and on the parametrization of quantum field, at any order of loop
expansion, is proportional to the effective equations of motion. As
an important consequence of this rule, the respective dependencies
in the one-loop divergences vanish on the classical mass shell.

The first feature implies that the general structure of divergences in
any loop order may be defined on the basis of power counting of the
Feynman diagrams, that means it can be reduced to the use of the
dimensional arguments. In this way we can, in most of the cases,
say that the given model of QG is renormalizable, non-renormalizable,
or super-renormalizable, and even establish the structure of possible
divergences in any loop order -- without making explicit calculations.
The remaining questions concern only the coefficients of the given
terms in the divergences.

The second feature is also important for practical purposes,
for the following two reasons. From one side, one can choose the
gauge fixing condition in such a way that makes the calculations
technically simpler.
And, on another hand, we can extract that part of the quantum
corrections which are gauge-fixing and parametrization invariant
and, in principle, consider this as a physical output of the loop
calculations.

The Chapter is organized as follows. In Sec.~\ref{secQG1.2} we
formulate the most relevant models of QG that may constitute a
sound basis of the perturbative treatment. Let us note that we only
slightly touch the nonlocal models which represent, nowadays, a
popular object of studies. The reason is that there is a special
Section devoted to the nonlocal theories and we do not like to
have repetitions. Thus, we mainly consider the QG based on GR
and on the different kinds of local polynomial models.
In Sec. \ref{IIQG-sec2.0}, one can find the bilinear expansions
of all metric-dependent terms that define the propagator of the
quantum metric.
In Sec.~\ref{secQG1.3}
we describe these propagators 
different models
of QG. The next Sec.~\ref{IIQG-sec1.4}  is briefly formulating the
the main statements about gauge-invariant renormalization in QG,
while the detailed discussion is postponed to the separate Chapter
of this Section.
Sec. \ref{secQG1.4} describes the power
counting and classification of the QG models to renormalizable,
non-renormalizable and super-renormalizable ones.
Sec.~\ref{IIQG-chap2} discusses the problem of ghosts in higher
derivative QG. This problem certainly represents the main difficulty
of all the QG program. We give a basic introduction and a
brief report on the existing results in this area.
In Sec.~\ref{secQG1.5.2} we describe the gauge-fixing and
parametrization dependencies of the one-loop effective action and
Sec.~\ref{secQG1.5} shows the detailed derivation of the one-loop
divergences in quantum GR, in the simplest parametrization and
gauge fixing.
Sec.~\ref{secQG1.5.3} discusses an interesting example of the
renormalization group applied to the quantum GR in a manner
that fits the effective approach to QG. There is a special Section
of the present Handbook, devoted to the subject of effective QG,
so we do not go too deep into the subject and only show this
particular example. Sec.~\ref{secQG1.xxx} gives a short review of
the one-loop calculations in other models of QG. Owing to the size
limitations, this part is very much incomplete and does not cover
the extensive literature on the subject, so it is included just as a
starting point for the interested readership.
Finally, in Sec.~\ref{secQG1.7} we draw our conclusions and
present final discussions of the current situation in perturbative QG.

In the rest of the Chapter, we use DeWitt notations, when the
covariant integral $\int d^4x\sqrt{-g}=\int_x$ may be assumed but not
written explicitly when this is obvious. We denote functional trace
$\Tr$ and determinant $\Det$, that includes the same covariant
integration over spacetime coordinates, and use pseudoeuclidean
notations such as $\sqrt{-g}$, even in case of using heat kernel
technique, that requires Euclidean metric. In these cases, we assume
the Wick rotation, regardless it may be a nontrivial issue in some
models of QG.
On top of this, the notations include the signature
\ $\eta_{\al\be} = \diag (+\,-\,-\,-)$,
the definition of the Riemann tensor
\beq
R^\la_{\,\,\,\tau\al\be} =
\pa_\al\,\Ga^\la_{\,\,\tau\beta}
- \pa_\be\,\Ga^\la_{\,\,\tau\al}
+ \Ga^\la_{\,\,\ga\al}\,\Ga^\ga_{\,\,\tau\be}
- \Ga^\la_{\,\,\ga\be}\,\Ga^\ga_{\,\,\tau\al} ,
\label{curva}
\eeq
Ricci tensor
${R^\al}_{\mu\al\nu} = R_{\mu\nu}$, and its trace
$R=R_{\mu\nu} g^{\mu\nu}$, that is Ricci scalar.

\section{Models of QG. General classification and gauge fixing}
\label{secQG1.2}

Our purpose is to construct the quantum theory of the metric field
$g_{\mu\nu}$. In GR, the metric is subject to the gauge transformation,
also called diffeomorphism, corresponding to the infinitesimal
coordinate transformation
\beq
x^\mu \to x^{\prime\,\mu} =  x^\mu + \xi^\mu,
\qquad
\xi^\mu= \xi^\mu(x).
\eeq
Let us start by deriving the diffeomorphism transformation for the
metric. Keeping only the terms of the first order in $\xi^\mu$ and
their derivatives, we get
\beq
g'_{\al\be}(x)
&=&
g'_{\al\be}(x')
- \frac{\pa g'_{\al\be}(x')}{\pa x'^\la}\,\xi^\la
= g'_{\al\be}(x')
- \pa_\la g_{\al\be}\,\xi^\la.
\nonumber
\eeq
Using the tensor transformation rule,
\beq
g'_{\al\be}(x')
&=&
\frac{\pa x^\rho }{\pa x'^\al}
\, \frac{\pa x^\si }{\pa x'^\be}\,
g_{\rho\si}(x)
\,=\,
\big( \de^\rho_\al - \pa_\al \xi^\rho\big)
\big( \de^\si_\be - \pa_\be \xi^\si\big)
g_{\rho\si}(x)
\nonumber
\\
&&
\quad
\,= \,\,
g_{\al\be}
- g_{\rho\be}\,\pa_\al \xi^\rho - g_{\al\rho}\,\pa_\be \xi^\rho
\eeq
and taking the two expressions together, we get
\beq
\de g_{\al\be}(x)
&\,=\,&
g'_{\al\be}(x)\,-\,g_{\al\be}(x)
\,=\, - g_{\la\be}(x)\,\pa_\al \xi^\la(x) - g_{\al\la}(x)\,\pa_\be \xi^\la(x)
\nonumber
\\
&&
\quad
-\,\, \pa_\la g_{\al\be}(x)\,\xi^\la(x)
\,\,= \,\,
 -\,\na_\al \xi_\be\,-\,\na_\be \xi_\al,
\label{qg1}
\eeq
that defines the generators of the gauge transformations
\beq
\de g_{\mu\nu}(x)
\,=\, R_{\mu\nu \,\la}(g)\,\xi^\la ,
\qquad
R_{\mu\nu \,\la}(g)
\,=\, -\,g_{\mu\la}\na_\nu\,-\,g_{\nu\la}\na_\mu
\label{qg3}
\eeq
We assume that any sort of the QG candidate action $S=S(g)$
possesses the symmetry under  (\ref{qg3}), i.e., satisfies the
Noether identity,
\beq
\frac{\de S}{\de g_{\mu\nu}}\,R_{\mu\nu \,\la}(g)\,=\, 0.
\label{qg2}
\eeq
In the next Chapter \cite{LavSha-HQG} (see also \cite{OUP}),
it is shown that QG is the gauge theory of the Yang-Mills type,
that means the algebra of the generators is closed off shell. This
feature represents the basis of the two statements \textit{i)} and
\textit{ii)}, mentioned in the Introduction.

In many situations, it is useful to parameterize the
metric as a perturbation over the Minkowski spacetime
\beq
&&
g_{\mu\nu} \,=\, \eta_{\mu\nu} + h_{\mu\nu},
\label{h}
\eeq
and use the following representation of $ h_{\mu\nu}$
(see \cite{Percacci} and Sec.~\ref{secQG1.3} for more details):
\beq
&&
h_{\mu\nu} \,\,=\,\,
{\bar h}^{\bot\bot}_{\mu\nu}
+ \pa_\mu \ep^\bot_\nu + \pa_\nu \ep^\bot_\mu
+  \pa_\mu \pa_\nu \ep
+ \frac14\,h\, \eta_{\mu\nu}.
\label{h_mn}
\eeq
In this expression, the tensor component (spin-2 mode) is traceless
and transverse, i.e., ${\bar h}^{\bot\bot}_{\mu\nu} \eta^{\mu\nu}=0$
and $\pa^\mu{\bar h}^{\bot\bot}_{\mu\nu}=0$. The irreducible vector
component  (spin-1 mode) satisfies the condition
$\pa_\mu \ep^{\bot\mu}=0$. There are also two scalar fields (or
modes) $\ep$ and $h$. The indices are raised and lowered with the
flat metric.

In the rest of this section, we describe the most popular actions
which are used for constructing the QG models. This description
includes the proper action and the details of the DeWitt-Fadeev-Popov
(or Fadeev-Popov) procedure required for the Lagrangian quantization
of the model. For the sake of generality, we perform the consideration
of these procedures using the background field method. The last
means that, instead of (\ref{h}), the expansion is performed around
an arbitrary background metric,
\beq
&&
g_{\mu\nu} \,\longrightarrow
\, g'_{\mu\nu} \,=\, g_{\mu\nu} + h_{\mu\nu}.
\label{bfm}
\eeq
The interested reader can find more details of the expansion
(\ref{bfm}) and useful exercises in the book \cite{OUP}.

\subsection{DeWitt-Faddeev-Popov method for QG}
\label{secQG1.1}

Let us briefly sketch the Faddeev-Popov (or DeWitt-Faddeev-Popov)
method for QG.
There is no critical difference with other gauge models, and we
shall use the notations close to the general ones. Namely, we shall
denote $g^i = g_{\mu\nu}$ and keep in mind that, depending on our
intentions, $g^i$ may be also used for $\,h_{\mu\nu}$, as defined in
(\ref{bfm}). Then the transformation rule and the generator from
Eq.~(\ref{qg3}) will be denoted as \
\beq
\de g^i = R^i_{\,\al} \xi^\al
\qquad
\mbox{and}
\qquad
R^i_{\,\al} = R_{\mu\nu\,\al}.
\label{cond}
\eeq
The last note about notations is that $g^i$ may be also used for
other parameterizations of the quantum metric\footnote{An example
is Eq.~(\ref{bgf}) below. }, and the formulas can be modified
accordingly.

The starting point is the naive expression for the functional
integral over the quantum metric
\beq
Z \,=\, \int  dg \,e^{iS(g)},
\label{Znaive}
\eeq
where we use $g \equiv g^i$ for the arguments of a functional
$S$ and the integration variable, to make formulas more readable.
We can generalize the last integral to the generating functional of
Green functions by replacing $S \to S + g J$ and assuming
that the source term $gJ = \int_x g^i J_i$ is diffeomorphism invariant.
As this replacement does not cause real changes, we shall work
with the formula (\ref{Znaive}).

The space of integration $g^i$ includes the \textit{orbits}, i.e. the
subspaces defined by the gauge transformations of the metric field
(\ref{cond}). Since the action remains constant over any such
subspace, it is cleat that each of these subspaces contributes
infinite to the integral, which is, therefore,  badly defined. Such
a divergence is similar to taking logarithm of determinant of a
degenerate matrix.  This divergence it is a direct consequence of
the gauge invariance and represents the problem solved by the
DeWitt-Faddeev-Popov method \cite{DeWitt-67,faddeev-popov}
in gravity and in the Yang-Mills theory (see also,  e.g.,
\cite{Popov-FI} for QG and further references in the
next Chapter \cite{LavSha-HQG}).

The first step is to insert the factor of unity in the integrand of
(\ref{Znaive}), in the form
\beq
1 \,=\, \De(g)\,\int d\xi^\al \,\,\de\big(\chi^\al(g) - l^\al \big).
\label{1}
\eeq
Here $l^\al$ is an arbitrary vector field and $\chi^\al(g)$ is the
gauge fixing condition. There can be different choices of this
condition but, in principle, one can think that  the surface
$\chi^\al(g)=l^\al$ (or, simply, $\chi^\al(g)=0$) crosses each
orbit in a unique ``point'', such that the degeneracy is removed.
Indeed, this is not the way  Faddeev-Popov method works,
as we will see.

The integral (\ref{1}) can be easily taken by making the change
of integration variables to $\chi^\al$,
\beq
\De^{-1}(g)
\,=\,
\,\int d\chi^\be \,\Det \Big( \frac{\de \xi^\al}{\de \chi^\be}\Big)
\,\de\big(\chi^\al(g) - l^\al \big).
\label{changevar}
\eeq
The Jacobian of this transformation is inverse to the matrix
\beq
\Big( \frac{\de \xi^\al}{\de \chi^\be}\Big)^{-1}
\,=\,\,
\frac{\de \chi_\al}{\de \xi_\be}
\,=\,
\frac{\de \chi_\al}{\de g^i}\,R^{i\,\be}
\,=\,
\frac{\de \chi_\al}{\de g_{\mu\nu}}\,R_{\mu\nu}^{\quad\,\be}
\,\,=\,\,
M_\al^{\,\,\be}.
\label{ghostMgen}
\eeq
It is important that the determinant of the matrix $M_\al^{\,\,\be}$
does not depend on $\xi$, but only on $g^i$ and the form
of the gauge fixing condition $\chi(g)$.
In this way, we arrive at the equivalent, albeit already
non-degenerate, form of (\ref{Znaive}),
\beq
Z \,=\, \int  dg \,\,
\,\de\big(\chi^\al(g) - l^\al \big) \,\Det \big( M_\al^{\,\,\be}\big)
\, \,e^{iS(g)}\,.
\label{Zintermed}
\eeq
To achieve a more useful form of this expression, we insert in
the integrand a unit in the form
\beq
1 \,=\, \big(\Det Y_{\al\be}\big)^{1/2}
\,\int d l^\al \,\exp\Big\{ \frac{i}{2}\,l^\al Y_{\al\be}l^\be \Big\}.
\label{1.1}
\eeq
Here $Y_{\al\be}$ is a new object identified as weight operator.
We will discuss several forms of this operator, adapted to different
models of QG, in what follows.

Taking the integral with the delta function, we get
\beq
Z \,=\, \int  dg \,\,
\,\big(\Det Y_{\al\be}\big)^{1/2}\,
\big(\Det M_\al^{\,\,\be}\big)
\, \,e^{iS + iS_{gf}},
\label{Z-1}
\eeq
where
\beq
S_{gf}
\,=\,
\,\frac12\,\chi^\al Y_{\al\be} \,\chi^\be.
\label{Sgf}
\eeq
In some situations, it is useful to deal with the modified
form of the operator,
\beq
\tilde{M}_{\al\be}
\,\,=\,\,
Y_{\al\la}\, M_\la^{\,\,\be},
\label{M-mod}
\eeq
providing an alternative form of (\ref{Z-1}), i.e.,
\beq
Z
&=&
\int  dg \,\,
\,\big(\Det \,Y_{\al\be}\big)^{-1/2}\,
\big(\Det \,\tilde{M}_\al^{\,\,\be}\big)
\, \,e^{iS + iS_{gf}}
\nn
\\
&&
\quad
= \,\, \int  dg \,d\bar{C} \,dC\,\,
\,\big(\Det Y_{\al\be}\big)^{-1/2}\,
\, \,e^{iS + iS_{gf}+ iS_{gh}},
\label{Z-2}
\eeq
where the action of the Faddeev-Popov ghosts has the form
\beq
S_{gh} \,=\,
\bar{C}^\al \,\tilde{M}_\al^{\,\,\be}\,C_\be\,.
\label{Sgh}
\eeq
It is clear that the ghost fields $\bar{C}^\al$ and $C_\be$
should be fermions (i.e., have odd Grassmann parity) to provide
the positive power of $\Det \tilde{M}_\al^{\,\,\be}$. One can
trade the functional determinant
$\big(\Det \,Y_{\al\be}\big)^{-1/2}$
for another (third) ghost. This ghost may have even or odd
Grassmann parity, respectively, for the versions with
$\,\tilde{M}_{\al\be}\,$ or $\, M_\la^{\,\,\be}$.

Thus, the total Faddeev-Popov action  \ $S_t = S + S_{gf}+ S_{gh}$
\ depends on the
choice of the gauge fixing condition $\chi^\al$ and the weight
operator (sometimes called weight function) $Y_{\al\be}$. In what
follows, we consider a few important examples of the choice of these
two objects in different models of QG. As
with the most of this review, the reader can find more detailed
discussion in \cite{OUP}.

\subsection{Quantum general relativity}
\label{QG-GR}

The first and most obvious candidate for the starting action of QG
is the Einstein-Hilbert functional
\beq
S_{EH}\,=\,-\,\frac{1}{\ka^2}\,\int d^4x \sqrt{-g}\,
\big( R + 2\La\big).
\label{EH}
\eeq
Here we denoted $\ka^2 = 16\pi G$, as it proves useful when we
consider the perturbative expansion. The reason is that, modifying
the expansion (\ref{bfm}) to
\beq
g_{\mu\nu} \quad \longrightarrow \quad
g^\prime_{\mu\nu} \,=\, g_{\mu\nu} + \ka h_{\mu\nu},
\label{qggr1}
\eeq
the bilinear in $h_{\mu\nu}$ terms are free from the parameter
$\ka$, while the higher orders in  $h_{\mu\nu}$ terms have
positive powers $\ka$. One of the consequences is that $\ka$
plays the role of the interaction constant in this QG model. In
quantum theory, $\ka$  turns out the loop expansion parameter.
Let us stress that this is, typically, not so for other QG models,
which may have other parameters of the loop expansion. Other
two relevant observations is that, since gravity is a non-polynomial
theory, the  parametrization (\ref{qggr1}) results in that the action
(\ref{EH}) has unbounded powers of  $\ka$. Furthermore, things
get more complicated in more general parameterizations but,
typically, $\ka$ remains the parameter of the loop expansion.

The flat limit requires a vanishing cosmological constant and
we get
\beq
g_{\mu\nu} \,=\, \eta_{\mu\nu} + \ka h_{\mu\nu}\,.
\label{qg-flat}
\eeq
As we shall see below, this means that the propagator of quantum
metric is free of $\ka$ by construction.

Instead of the expansion of the metric $g_{\mu\nu}$, one can
start from the expansion of the inverse metric $g^{\mu\nu}$, or
even a more general parametrization of the quantum metric. Later
on, we consider the version which is the most general for the
one-loop calculations and has various arbitrary parameters.

The Faddeev-Popov procedure requires introducing the gauge
fixing term $S_{gf}$. Such a term is called to make the highest
derivative in $h_{\mu\nu}$ terms of the action $S+S_{gf}$
non-degenerate. Then one can obtain the propagator of this field
or, e.g., apply the heat-kernel method and Schwinger-DeWitt
technique for calculating the divergences in a covariant form.
Since the theory  (\ref{EH}) has at most second derivatives, we
can choose the weight operator proportional to the background
metric and arrive at the gauge-fixing term in the form
\beq
S_{gf}\,=\,\frac{1}{\al}\,\int d^4x \sqrt{-g}\,
\chi_\mu\,\chi^\mu ,
\qquad
\mbox{where}
\qquad
\chi_\mu = \na_\nu h^\nu_{\,\,\mu}- \be \na_\mu h
\label{qg-EH-gf}
\eeq
and $\al$ and $\be$ are the gauge-fixing parameters. The dependence
on the choice of the gauge fixing and on the parametrization  of
quantum metric, represents an important part of the QG development.

Why the action (\ref{EH}) requires the simplest weight functional
in the case of QG? The reason is that the Lagrangian of GR (\ref{EH})
has at most two derivatives of the quantum metric $h_{\mu\nu}$ in
the action. Thus, a ``correct'' (or, better to say, appropriate)
way of breaking down the degeneracy in the total action requires
that the gauge fixing term $S_{gf}$ also has two derivatives. Since
each of $\chi_\mu$ in Eq.~(\ref{qg-EH-gf})
is linear in derivatives, we are forced to implement the choice
$Y_{\mu\nu} = \mbox{const} \times g_{\mu\nu}$ in this case. With some
adjustments, the same logic will be used in all models of QG which
will be discussed below (and even those which will not be).

The action of ghosts is constructed in a standard way, as
\beq
S_{gh}\,=\,\int d^4x \sqrt{-g}\,\,\,
\bar{C}^\al\, M_\al^{\,\,\,\be}\,C_\be,
\label{qg-EH-gh}
\eeq
where the operator $M$ is the variation of gauge fixing
condition with respect to the transformation function,
\beq
M_\al^{\,\,\,\be}
\,=\,
\frac{\de \chi_\al}{\de \xi_\be}
\,=\,
\frac{\de \chi_\al}{\de h_{\mu\nu}}\,
R_{\mu\nu }^{\quad\,\be}\,.
\label{ghostM}
\eeq
It is important that the ghost fields in (\ref{qg-EH-gh}) satisfy
the second-order equations. This means, the propagators of both
gravitational field  $h_{\mu\nu}$ and ghosts have the same type
of the UV behavior $1/k^2$. In the next models, we shall see that
providing the homogeneity of the propagators of different modes
of the metric (\ref{h_mn}) and of the ghosts may require some extra
efforts. And such a homogeneity worth these efforts, as without it,
the quantum theory gains a lot of artificial complications.

\subsection{Fourth derivative gravity}
\label{QG-4d}

The next model of common interest is based on the fourth-derivatives
action. If the previous choice, namely, the Einstein-Hilbert action
of GR, is strongly motivated by the success of Einstein's classical
gravitational theory, the fourth-derivative model is motivated by
the consistency conditions of semiclassical gravity. It is known from
the early paper by Utiyama and DeWitt \cite{UtDW} (see also the
aforementioned books \cite{birdav,book,ParToms,OUP}) that if
the matter fields are quantum, the action of vacuum (i.e., the
gravitational action) of renormalizable theory has to include both
Einstein-Hilbert action (\ref{EH}) and the covariant local
fourth-derivative terms
\beq
S_{HD}\,=\,\int d^4x\sqrt{-g}\,
\Big\{a_1 R_{\mu\nu\al\be}^2
+ a_2 R_{\mu\nu}^2
+ a_3 R^2
+ a_4 \cx R
\Big\}.
\label{HD}
\eeq

In QG, it is more useful to write this action in another basis,
including the square of the Weyl tensor and the integrand of
the Gauss-Bonnet topological term
\beq
&&
C^2
\, = \,
C_{\mu\nu\al\be}C^{\mu\nu\al\be}
\, = \,
R^2_{\mu\nu\al\be} - 2R_{\mu\nu}^2 + \frac13 R^2,
\nn
\\
&&
E_4\, = \,
R^2_{\mu\nu\al\be} - 4R_{\mu\nu}^2 +  R^2.
\label{C2E4}
\eeq

In this basis, the fourth-derivative action has the form
\beq
S_{HD}\,=\,-\,\int d^4x\sqrt{-g}\,
\Big\{\frac{1}{2\la}C^2 \,-\,\frac{1}{\rho}E_4
\,+\,\frac{1}{\xi} R^2 \,+\,\tau\square R
+ \frac{1}{\ka^2} (R - 2 \La )\Big\},
\label{action4der}
\eeq
Sometimes other notations for the couplings $\rho$ and $\xi$ are
used, e.g., $\th = \la/\rho$ and $\om = - 3\la/\xi$, or the ones used
in the special chapter \cite{Nobu-HD} about the
one-loop calculations and renormalization group flows in the
theory (\ref{HD}). The important part is the positive sign of the
coupling $\la$, as it is required by the positivity of the
energy of the massless tensor mode (graviton) in the high energy
region \cite{Stelle77,stelle78} (see also an alternative treatment
of the same problem in \cite{OUP}).

Introducing the gauge-fixing term in the fourth-derivative theory
of QG is a more complicated task compared to the quantum GR.
Since we are interested to maintain homogeneity of the propagator,
the gauge-fixing term should have four derivatives of the quantum
metric and hence we introduce the expression for the gauge-fixing
action
\beq
S_{gf}\,=\, \frac{1}{2}
\int d^4x\sqrt{-g}\,\,\chi^\mu\,Y_{\mu\nu}\,\chi^\nu\,,
\label{Sgf-QG}
\eeq
where the gauge condition $\chi^\mu$ is still defined by the formula
(\ref{qg-EH-gf}), but the new weight function $Y_{\mu\nu}$ should
be a non-degenerate operator of the second order in derivatives. In
the framework of background field method, its most general form is
\beq
Y_{\mu\nu} & = & \frac{1}{\al}\,\Big( g_{\mu\nu}\Box +
\ga\na_\mu\na_\nu - \na_\nu\na_\mu + p_1R_{\mu\nu}
+p_2R\,g_{\mu\nu}\Big)\,,
\label{weight-4der}
\eeq
where the non-degeneracy requires $\ga \neq 0$. Compared to the
quantum GR, the new gauge-fixing condition depends on a larger
number of arbitrary gauge-fixing parameters, i.e., on
$\,\al_i\,=\,(\al,\,\be,\,\ga,\,p_1,\,p_2)$, where $\be$ comes from
Eq.~(\ref{qg-EH-gf}). In the case of the flat background metric,
$Y_{\mu\nu}$ has only two arbitrary parameters $\al$ and $\ga$.

With the definition (\ref{Sgf-QG}) and  (\ref{weight-4der}),
all the modes of the quantum metric in (\ref{h_mn}) have the
same leading UV behaviour of the propagator,
$G^{-1}_i(k) \propto k^4$.
As we shall see in the forthcoming sections, the power counting
is greatly simplified if the ghost action has the same number of
derivatives as the action of the $h_{\mu\nu}$ field. It is clear that
the problem can be solved by introducing a modified ghost action
(\ref{Sgh}), namely
\beq
&&
S_{gh}\,=\,\int d^4x\sqrt{-g}\,\, {\bar C}^\al
\,{\tilde M}_\al^{\,\,\,\be}
\,C_\be\,,
\qquad
\mbox{where}
\qquad
{\tilde M}_\al^{\,\,\,\be}\,=\,
M_\al^{\,\,\,\la}(g)\,Y_\la^{\,\,\,\be}(g).
\mbox{\qquad}
\label{Mghost-mod}
\eeq

The functional integral is defined by the general expression
(\ref{Z-2}) with the specific choice (\ref{weight-4der}) of the
weight operator. Since both $\,M_\al^{\,\,\la}\,$ and
$\,Y_\la^{\,\,\be}\,$ are second order operators, the propagator
of the Faddeev-Popov ghosts behaves like $\,k^{-4}$
in the UV, exactly as the propagator of gravitational perturbations.
This feature proves useful for evaluating the power counting of
the Feynman diagrams in this theory.

\subsection{Quantum gravity models polynomial in derivatives }
\label{IIQG-sec1.2.3}

The previous two examples of QG models are minimal versions.
In particular, GR fits all observational and experimental tests for
a classical gravity \cite{Will-book} and, in this sense, can be
regarded as a reference theory. On the contrary, there is not a
single experimental test for the fourth-derivative model of gravity
but, from another side, fourth derivative terms are required for the
renormalizability of semiclassical gravity. And the same model
guarantees also the renormalizability of QG \cite{Stelle77}. On
another hand, the fourth derivative theory has serious problems
related to nonphysical ghosts\footnote{Which have nothing to do
with the Faddeev-Popov ghosts, the traditional use of the same word
here is a mere coincidence.} and stability, as we shall discuss
below. In this situation, one may look beyond the minimal theories
and it is natural to try the models with more than four derivatives.
Then, the number of derivatives may be finite or infinite. In the
first case we meet the polynomial in derivatives models of QG,
suggested in \cite{highderi}.

To construct the polynomial models, we impose the condition that
the highest-derivative terms in the action should be homogeneous
in the derivatives. In the next section, we shall see that this
homogeneity may provide the superrenormalizability of the theory.
The action of the theory has the form
\beq
S_{N}
&=&
\int \!d^4x \sqrt{-g}\,
\Big\{
\te_{N,R} R \, \square^{N} R
+ \te_{N,C} C \, \square^{N} C
 + \te_{N,\mathrm{GB}}\mathrm{GB}_{N}
\nonumber
\\
&& 
+  \, \te_{N-1,R}R \, \square^{N-1}R
+\te_{N-1,C}C \, \square^{N-1}C
+ \te_{N-1,\mathrm{GB}}\mathrm{GB}_{N-1}
\, + \,\,\, ... \,\,\,
\nonumber
\\
&& 
+  \,  \te_{0,R}R^2
 +\te_{0,C}C^2
+ \te_{0,\mathrm{GB}}\mathrm{GB}_{0}
 +\te_{\rm EH}R
 +\te_{\rm cc} + {\mathcal O}(R_{\dots}^3)
 \Big\}\,,
 \mbox{\qquad}
 \label{gaction}
\eeq
where $N=1,2,\dots$ and all $\te$'s are arbitrary parameters of the
action.  It is assumed that both $\te_{N,R}$ and $\te_{N,C}$ are
non-zero and that the maximal power of metric derivatives in the
terms ${\mathcal O}(R_{\dots}^3)$ is $2N+4$, i.e., is not higher
than of the terms of the second order in curvatures, i.e.,
${\mathcal O}(R_{\dots}^2)$.

Furthermore,  there are the squares of the Weyl
tensor with extra factors of $\,\cx$,
\beq
C \, \square^n\,C
\,=\,
C_{\mu\nu\al\be}\, \square^n\,C^{\mu\nu\al\be}
\,=\,
R_{\mu\nu\al\be} \square^{n} R^{\mu\nu\al\be}
- 2 R_{\mu\nu} \square^{n} R^{\mu\nu}
+ \frac13\,R \square^{n} R.
\mbox{\qquad}\mbox{\quad},
\label{C2n}
\eeq
Similarly, using integrations by parts and the Bianchi identities,
the generalized Gauss-Bonnet invariants can be shown to have
the property
\beq
\mathrm{GB}_{n}
\,=\,
R_{\mu\nu\al\be} \square^{n} R^{\mu\nu\al\be}
- 4 R_{\mu\nu} \square^{n} R^{\mu\nu}
+ R \square^{n} R \,=\, {\mathcal O}(R_{\dots}^3).
\label{GBn}
\eeq
This term is not topological for $\,n \geq 1$, but it contributes
only to the third- and higher-order terms in the curvature
tensor, and to the total derivatives.
Thus, it may affect the vertices but not the propagators of QG,
as we shall explicitly check out in what follows.
On a flat background, the ${\mathcal O}(R_{\dots}^2)$ terms
may contribute to the propagator of the gravitational perturbation,
while ${\mathcal O}(R_{\dots}^3)$ terms affect only the vertices.
In the action (\ref{gaction}), the ${\mathcal O}(R_{\dots}^2)$'s
are given in the basis of Weyl-squared and $R$-squared terms.
As will be shown below, the  Weyl-squared terms affect the
propagation of the tensor mode ${\bar h}^{\bot\bot}_{\mu\nu}$,
and the $R$-squared $R\cdot R$ terms affect the propagation of
the scalar modes.  Using the higher-derivative actions in the form
(\ref{gaction}), we separate the propagators of the tensor and
scalar modes at the level of the action.

To provide the homogeneity in derivatives for the propagator
of all modes (\ref{h_mn}), the gauge-fixing terms should have the
same highest power in derivatives as the main action. Since the
gauge fixing conditions $\chi_\mu$ are always chosen in the form
(\ref{qg-EH-gf}), the $\mathcal{O}\big(p^{-4-2N}\big)$ propagator
of the quantum metric requires the weight function to be
\beq
Y_{\mu\nu}
\,=\,
-\,\frac{1}{\alpha}\big(g_{\mu\nu}\square
+ \gamma\nabla_{\mu}\nabla_{\nu}
- \nabla_{\nu}\nabla_{\mu}\big)\square^{N+1}.
\label{weight-N}
\eeq
One can add here many terms with lower order of derivatives, but
this does not change the UV behaviour of the propagator.

To provide the same power of derivatives in the ghost sector,
as in the quadratic in curvature action, one can redefine the ghost
action as (\ref{Mghost-mod}), this time with the weight operator
(\ref{weight-N}).

Furthermore, one can write the action (\ref{gaction}) in
an alternative form,
\beq
S_N
& = &
\int d^4x
\sqrt{-g}\,
\Big\{
\frac12\,
C_{\mu\nu\al\be} P_1(\square) C^{\mu\nu\al\be}
\,+\,
\frac12\,R P_2(\square) R
\nn
\\
&&
\quad + \,\,\,\te_{\rm EH}R
\,+\, \te_{\rm cc}
\,+\, {\mathcal O}(R_{\dots}^3)
\Big\},
\mbox{\qquad}\mbox{\quad}
\label{poliaction}
\eeq
where $ P_{1,2}(x)$ are polynomials of the same order $N$
and the terms ${\mathcal O}(R_{\dots}^3)$ have at most
$4+2N$ derivatives of the metric.
One can make further generalization of (\ref{poliaction}), by
trading the polynomials to the infinite series of $\cx$. The
discussion of these theories can be found in the
corresponding Section of the present Handbook. Let us just
quote the expression for the general (i.e., polynomial or
non-polynomial) action
\beq
S_{gen}
&=&
\int d^4 x \sqrt{-g} \,
\Big\{ - \frac{1}{\ka^2} \,(R+2\La)
\,+\, \frac12\,C_{\mu\nu\al\be} \,\Phi(\Box)\,C_{\mu\nu\al\be}
\nn
\\
&&
\quad
+ \,\,
\frac12\,R \, \Psi (\Box)\, R
\,+\, {\mathcal O}(R_{\dots}^3)\Big\}\,.
\label{act Phi}
\eeq
To complete the bilinear in curvature part of the action we can add
the third higher-derivative term, which boils down to the Gauss-Bonnet
topological term for a constant form factor $\Om$,
\beq
S_{GB}
&=&
\frac12 \int d^4 x \sqrt{-g} \,
\Big\{ R_{\mu\nu\al\be} \,\Om(\Box)\,R_{\mu\nu\al\be}
\,-\, 4R_{\mu\nu} \,\Om(\Box)\,R_{\mu\nu}
\,+\, R\,\Om(\Box)\,R\Big\}.
\qquad
\label{GBterm}
\eeq
Despite this term is equivalent to ${\mathcal O}(R_{\dots}^3)$
in (\ref{poliaction}), it makes sense to verify this feature, at
least at the level of the propagator.
The homogeneity of the propagator of all modes of the metric
perturbations (\ref{h_mn}), requires the functions $\Phi(x)$ and
$\Psi (x)$ to have analogous behavior in the UV. This can be
achieved by requiring that
\beq
\lim\limits_{x\to \infty} \,\frac{\Psi (x)}{\Phi(x)} \,\,=\,\, C,
\label{cond-nonloc}
\eeq
with $\,C\,$ being a non-zero constant. For the polynomial
functions $ P_{1,2}(x)$ of the same order, in (\ref{poliaction}),
the last condition is guaranteed. In the case of the non-polynomial
functions, the simplest useful choice is
\beq
\Phi(x) = - \frac{1}{\ka^2\,x}\,\big( e^{\al_1x}-1 \big)
\quad
\mbox{and}
\quad
\Psi (x) = -  \frac{C}{\ka^2\,x}\,e^{\al_2x},
\label{PhiPsi}
\eeq
such that condition (\ref{cond-nonloc}) reduces to  $\al_1=\al_2$.

The gauge-fixing term in theory (\ref{act Phi}) that provides the
homogeneity of the propagators, is of the standard form
(\ref{Sgf-QG}), but  requires the special weight
operator
\beq
Y_{\mu\nu}
\,\,= \,\,
\big(g_{\mu\nu}\square - \gamma\nabla_{\mu}\nabla_{\nu}
+ p_1 R_{\mu\nu} + p_2 R g_{\mu\nu}\big)\, W(\cx).
\label{weight-nonloc}
\eeq
In many cases, it is sufficient to consider $W(\cx)\propto\Phi(\cx)$,
but we shall keep this function arbitrary for generality, until some
point.

Finally, the homogeneity in the  momentum at the UV for the
quantum metric and for the Faddeev-Popov ghosts, can be
achieved by the standard replacement (\ref{Mghost-mod}).

\section{Bilinear forms and linear approximation}
\label{IIQG-sec2.0}

The analysis of propagators in different models of quantum gravity
requires the bilinear expansions of the relevant quantities depending
on the curvature tensor, on a flat metric background. On the other
hand, similar expressions with an arbitrary background metric are
useful for the one-loop calculations in the background field method.
Thus, we consider a general case and assume the expansion (\ref{bfm}),
i.e. $g_{\mu\nu}\,\longrightarrow\,g^\prime_{\mu\nu}
\,=\,g_{\mu\nu}+h_{\mu\nu}$.

Let us refer the interested reader to the book \cite{OUP} for
technical details and only give the following list of basic expansions:
\beq
&&
{g'}^{\mu\nu} = g^{\mu\nu}-h^{\mu\nu}+h^\mu_{\,\,\la} h^{\nu\la}
-h^\mu_{\,\,\la} h^\la_{\,\,\tau} h^{\nu\tau} + ..
\nn
\\
&&
\sqrt{- g^\prime} = \sqrt{- g} \, \Big( 1 + \frac12\,h
+\frac18\,h^2 - \frac14\,h_{\mu\nu}h^{\mu\nu} + \,...\Big),
\nn 
\\
&&
{\Ga^\prime}^\al_{\,\,\be\ga} \,=\,
\Ga^\al_{\,\,\be\ga} + \de \Ga^\al_{\,\,\be\ga}\,,
\label{action 7}
\eeq
where
\beq
&&
\de \Ga^\al_{\,\,\be\ga}
\,=\,
\frac12
\big( g^{\al\la}-h^{\al\la}+h^\al_\ka h^{\la\ka}
-h^\al_\ka h^\ka_\tau h^{\tau\la} + ...\big)\,
\big( \na_\be h_{\ga\la}+\na_\ga h_{\be\la}
- \na_\la h_{\ga\be} \big).
\nonumber
\eeq
In these formulas, the Greek indices are lowered and raised with
the background metric $g_{\mu\nu}$ and its inverse $g^{\mu\nu}$.
One has to remember that the variation $\de {\Ga}^\la_{\mu\nu}$
is a tensor, and, therefore, can be a subject to covariant
differentiation.

The first  two orders of expansion of the Riemann
tensor have the form
\beq
R^{\prime \al}_{\,\,\,\cdot\,\be\mu\nu}
&=& R^{\al}_{\,\cdot\,\be\mu\nu}
 + \de R^{\al}_{\,\cdot\,\be\mu\nu},
\quad
\mbox{where }
\quad
\de R^{\al}_{\,\cdot\,\be\mu\nu}
= R^{(1)\al}_{\quad\,\cdot\,\be\mu\nu}
+ R^{(2)\al}_{\quad\,\cdot\,\be\mu\nu},
\mbox{\qquad}
\label{Riem delta}
\\
R^{(1)\al}_{\quad\,\cdot\,\be\mu\nu}
&=& \frac12 \Big(
\na_\mu\na_\be h^\al_{\,\nu}
- \na_\nu\na_\be h^\al_{\,\mu}
+ \na_\nu\na^\al h_{\mu\be}
- \, \na_\mu\na^\al h_{\nu\be}
\nn
\\
&&
\,\,\,
+ \,R^\al_{\,\,\tau\mu\nu} h^\tau_{\,\be}
- R^\tau_{\,\,\be\mu\nu} h_{\tau\al}\Big),
\mbox{\qquad}
\mbox{\,\,}
\nn 
\\
R^{(2)\al}_{\quad\,\cdot\,\be\mu\nu}
&=& \frac12 h^{\al\la}
\big\{
\na_\mu\na_\la h_{\nu\be}
- \na_\nu\na_\la h_{\mu\be}
+ \na_\nu\na_\be h_{\mu\la}
- \na_\mu\na_\be h_{\nu\la}
\nn
\\
&&
\,\,\,
+ \, \big[\na_\nu,\na_\mu\big] h_{\be\la}\big\}
+ \frac14  \Big\{
(\na_\mu h^{\al\la})
(\na_\la h_{\nu\be}- \na_\be h_{\nu\la} - \na_\nu h_{\la\be})
\nn
\\
&&
\,\,\,
- \,(\na_\be h^\la_{\,\mu}+ \na_\mu h^\la_{\,\be})
(\na_\la h^\al_{\,\nu} - \na^\al h_{\la\nu})
+ (\na^\la h_{\nu\be})(\na^\al h_{\mu\la} - \na_\la h^\al_{\,\mu})
\nonumber
\\
&&
\,\,\,
- \,(\na_\nu h^{\al\la})
(\na_\la h_{\mu\be}- \na_\be h_{\mu\la} - \na_\mu h_{\la\be})
\,- \,(\na^\la h_{\mu\be})(\na^\al h_{\nu\la} - \na_\la h^\al_{\,\nu})
\nonumber
\\
&&
\,\,\,
+\, (\na_\be h^\la_{\,\nu}+\na_\nu h^\la_{\,\be})
(\na_\la h^\al_{\,\mu} - \na^\al h_{\la\mu})
\Big\}.
\nn  
\eeq
Here and in what follows, points indicate the positions of the raised
indices. Similar formulas for the Ricci tensor and scalar curvature
are
\beq
&&
{R'}_{\be\nu}
= R_{\mu\nu} + \de{R}_{\mu\nu},
\quad
\mbox{where }
\quad
\de {R}_{\mu\nu}
 = R^{(1)}_{\mu\nu} + R^{(2)}_{\mu\nu},
\nn 
\\
&&
R^{(1)}_{\mu\nu}
= \frac12 \big(
\na_\la\na_\mu h^\la_{\,\nu} + \na_\la\na_\nu h^\la_{\,\mu}
- \na_\mu\na_\mu h -  \cx h_{\mu\nu}\big),
\\
&&
R^{(2)}_{\mu\nu}
=
\,\frac12 h^{\al\be}
\big(\na_\al\na_\be h_{\mu\nu}
+ \na_\mu\na_\nu h_{\al\be}
- \na_\al\na_\mu h_{\be\nu}
- \na_\al\na_\nu h_{\be\mu}\big)
\nonumber
\\
&&
\quad
+\,\frac12 \big(\na_\al h_{\mu\be}\big) \big(\na^\al h_\nu^{\,\,\be}
- \na^\be h_\nu^{\,\,\al}\big)
+ \frac14(\na_\mu h_{\al\be})(\na_\nu h^{\al\be})
\nonumber
\\
&&
\quad
+ \, \frac14\, \big(2\na_\be h^{\al\be} - \na^\al h\big)
\big( \na_\al h^{\mu\nu}- \na_\mu h^{\al\nu}-\na_\nu h^{\al\mu}\big)
\nn 
\\
&&
\mbox{and}
\qquad
R' \,=\,R\,+\,\de R,
\quad
\mbox{where }
\quad
\de R \,=\, R^{(1)}+ R^{(2)},
\label{R-expand}
\\
&&
R^{(1)}
\,= \,\na_\mu \na_\nu h^{\mu\nu} - \cx h - R_{\mu\nu} h^{\mu\nu} ,
\nn 
\\
&&
R^{(2)}
\,=\,
h^{\al\be} \Big(\na_\al\na_\be h
+  \cx h_{\al\be}
- \na_\al\na_\mu h^\mu_{\,\,\be}
-  \na_\mu  \na_\al h^\mu_{\,\,\be}\Big)
- \,\frac14(\na_\al h)(\na^\al h)
\nonumber
\\
&&
\quad
+ \frac14 \big(\na_\mu h_{\al\be}\big)
\big(3\na^\mu h^{\al\be} - 2\na^\al h^{\mu\be}\big)
\,+\, (\na_\al h^{\al\be})(\na_\be h - \na_\mu h^\mu_{\,\,\be})
+ R_{\mu\nu} h^\mu_{\,\,\al} h^{\nu\al}.
\nn 
\eeq

Using the expressions listed above, one can easily get the
expansions of the terms in the action, up the second order
in $\,h_{\mu\nu}$. The results can be written for the terms
in the four derivative action (\ref{action4der}), but it is easy
to show how, in the particular case of a flat metric, these
expansions can be mapped to the more general action
(\ref{act Phi}) with an arbitrary (finite or even infinite)
number of derivatives.

The first expansion has the form
\beq
\label{EH-expand}
&&
\Big(\int d^4 x \sqrt{-g'}\big[ R' + 2\La\big]\Big)^{(2)}
\,=\,
\frac14 \int d^4 x \sqrt{-g}\,h^{\mu\nu}\Big[
\de_{\mu\nu , \al\be}\cx
- g_{\mu\nu} g_{\al\be}\cx
\nonumber
\\
&&
\quad
- \,2 g_{\mu\al} \na_\nu \na_\be
+ \big(g_{\mu\nu} \na_\al \na_\be - g_{\al\be} \na_\mu \na_\nu \big)
- \big(g_{\mu\nu} R_{\al\be} - g_{\al\be} R_{\mu\nu} \big)
\nonumber
\\
&&
\quad
+ \,2R_{\mu\al\nu\be}
\,- \,\big(R + 2 \La \big)
\Big( \de_{\mu\nu , \al\be} - \frac12 g_{\mu\nu}g_{\al\be}\Big)
\Big] h^{\al\be}\,,
\nonumber
\eeq
where we used the DeWitt notation for the
unit matrix in the symmetric tensors space
\beq
\de_{\mu\nu,\al\be} \,=\,\frac12\,\big(
 g_{\mu\al}  g_{\nu\be} + g_{\nu\al}g_{\mu\be} \big)
\label{delta_abmn}
\eeq
and the short notations that assume the symmetrization, e.g.,
\beq
g_{\mu\al} \na_\nu \na_\be
\,\,
&\longrightarrow&
\,\,
\frac14\big(g_{\mu\al} \na_\nu \na_\be
+ g_{\nu\al}  \na_\mu \na_\be
+ g_{\mu\be}\na_\nu \na_\al
+ g_{\nu\be} \na_\mu \na_\al\big).
\label{sim rest}
\eeq

For the sake of brevity, the remaining expansions will be given
only for a flat background. The complete expressions can be
found, e.g., in \cite{OUP}.
The simplest of the remaining relevant bilinear expansions is
\beq
\Big( \int d^4 x \sqrt{-g'}{R'}^2\Big)^{(2)}_{\rm flat}
&=&
\int d^4 x \,\,\,h^{\mu\nu}\Big[
\eta_{\al\be}\eta_{\mu\nu}\cx^2
- \eta_{\mu\nu}\cx \pa_\al \pa_\be
\nonumber
\\
&&
- \,\, \eta_{\al\be}\pa_\mu \pa_\nu \cx
\,+ \,\pa_\mu \pa_\nu \pa_\al \pa_\be   \Big] h^{\al\be}.
\label{R2-expand}
\eeq
One can note the absence of the term
$\de_{\mu\nu , \al\be}\cx^2$ in this expression.
As a result, the $R^2$ term
does not affect the propagation of the spin-2 mode
${\bar h}^{\bot\bot}_{\mu\nu}$ of the metric perturbation
(\ref{h_mn}).

The next expansions are the one for the square of the Riemann tensor,
\beq
\Big( \int d^4 x \sqrt{-g'}{R'}_{\mu\nu\al\be}^2\Big)^{(2)}_{\rm flat}
&\,\,=\,\,&
\int d^4 x \,\,\,h^{\mu\nu}\Big[
\de_{\mu\nu , \al\be}\cx^2
+ \pa_\al  \pa_\be \pa_\mu \pa_\nu
\nonumber
\\
&&
-\,\, 2 \eta_{\nu\be} \cx \pa_\mu \pa_\al \Big] h^{\al\be}
\label{Rim2-expand}
\eeq
and for the square of the Ricci tensor,
\beq
&&
\Big( \int d^4 x \sqrt{-g'}{R'}_{\mu\nu}^2\Big)^{(2)}_{\rm flat}
\,=\,
\frac12
\int d^4 x \,\,\,h^{\mu\nu}\Big[
\frac12\,\big(
\de_{\mu\nu , \al\be} + \eta_{\mu\nu}\eta_{\al\be}\big) \cx^2
- \, \eta_{\nu\be} \cx\na_\al \na_\mu
\nn
\\
&&
\qquad
+ \,\na_\al  \na_\mu \na_\be \na_\nu
- \,\frac12 \eta_{\mu\nu} \cx \na_\al \na_\be
- \frac12\, \eta_{\al\be} \cx \na_\mu \na_\nu
\Big] h^{\al\be}\,.
\label{Ricci2-expand} 
\eeq
The last two expansions (\ref{Rim2-expand}) and
(\ref{Ricci2-expand}) possess the $\de_{\mu\nu , \al\be}\cx^2$ terms.
This means, these two terms contribute to the flat-space propagator
of the transverse and traceless mode of the gravitational perturbation
${\bar h}^{\bot\bot}_{\mu\nu}$ in the representation (\ref{h_mn}).

One can note that it would be impossible to have only one of the
terms (\ref{Rim2-expand}) and (\ref{Ricci2-expand}) contributing
to the propagation of the spin-2 mode, because (\ref{R2-expand})
does not contribute to this mode and the linear combination of the
three
terms (\ref{C2E4}) form a topological invariant $E_4$.

\section{Gravitational waves, quantization, and gravitons}
\label{IIQG-sec-GW}

The gravitational wave is a dynamical classical solution of
Einstein's GR without matter sources. This term can be also used for
the solutions of the same sort in modified gravity models. However,
since in these models the additional modes are typically massive and,
therefore, do not propagate for a long distances, it is most common
to attribute the notion of a gravitational wave to the solutions in
GR\footnote{The topics related to the models of massive gravity
are left beyond the present Handbook. The main reason is that, in
these models, quantum aspects do not play an important role.}.
Nowadays, gravitational waves represent
one of the most successful parts of gravitational physics, both
experimental and theoretical. However, in this short section we
present only a brief survey of the basic facts concerning the
gravitational waves on a flat background in GR.

\subsection{Gravitational waves in a weak-gravity regime}
\label{IIQG-sec-GW-GR}

We start with an action of gravity with $\La = 0$ and use the
bilinear expansion (\ref{EH-expand}) on the flat background
metric $g_{\mu\nu}=\eta_{\mu\nu}$. To discuss the emission
of the gravitational wave (in the simplest case), we also add the
action of matter and consider it approximation that provides a
linear equation for $h_{\mu\nu}$. In this way, we obtain the
action of GR in the linearized regime,
\beq
S_{total}^{(lin)}
&=&
- \frac{1}{32 \pi G}
\int d^4 x \,\,\,h^{\mu\nu}\Big\{
\frac12\,\de_{\mu\nu , \al\be}\cx
- \frac12\,\eta_{\mu\nu} \eta_{\al\be}\cx
- \eta_{\mu\al} \pa_\nu \pa_\be
\mbox{\qquad}
\nonumber
\\
&&
\quad
+ \,\frac12
\big(\eta_{\mu\nu} \pa_\al \pa_\be - \eta_{\al\be} \pa_\mu \pa_\nu \big)
\Big\} h^{\al\be}
\,\,-\,\,\frac12   \int d^4 x \,h^{\mu\nu}T_{\mu\nu},
\label{GR-lin}
\eeq
where $\cx = \eta^{\mu\nu}\pa_\mu \pa_\nu$ and
$T_{\mu\nu}$ is the energy-momentum tensor of matter in flat
spacetime background. The equation for metric perturbations has
the form
\beq
&&
\Big\{
\de_{\mu\nu , \al\be}\cx
- \eta_{\mu\nu} \eta_{\al\be}\cx
- 2 \eta_{\mu\al} \pa_\nu \pa_\be
+ (\eta_{\mu\nu} \pa_\al \pa_\be - \eta_{\al\be} \pa_\mu \pa_\nu
\Big\} h^{\al\be}
\,=\, 16 \pi G T_{\mu\nu}.
\nonumber
\\
&&
\label{Eq-lin-T}
\eeq

Multiplying both sides of Eq.~(\ref{Eq-lin-T}) by the matrix
\beq
K^{\mu\nu , \rho\si} \,=\, \de^{\mu\nu , \rho\si}
\,-\, \frac12\, \eta^{\mu\nu} \eta^{\rho\si},
\label{mat inv}
\eeq
we arrive at the equation for the modified stress tensor,
\ $S_{\mu\nu} = T_{\mu\nu}
- \frac12 T^{\,\la}_{\la}\, g_{\mu\nu}$,
\beq
\pa_\la\pa_\nu h^\la_{\,\mu}
+ \pa_\la\pa_\nu h^\la_{\,\mu}
- \cx h_{\mu\nu}
- \pa_\mu \pa_\mu h
\,=\, 16 \pi G \, S_{\mu\nu}\,.
\label{Eq-lin-S}
\eeq
Here $h =  h_{\mu\nu}\eta^{\mu\nu}$. The last equation
describes both propagation and emission of the gravitational
waves in the linear approximation. This equation has to be
supplemented by the gauge transformation (\ref{qg1})
$\de h_{\mu\nu} \,=\,
-\,\pa_\mu \xi_\nu\,-\,\pa_\nu \xi_\mu$,
and requires imposing the gauge-fixing condition, e.g.,
using the de Donder (also called Fock-de Donder) condition
\beq
\pa_\mu h^\mu_{\,\,\nu} \,-\, \frac12 \,\pa_\nu h \,=\,0.
\label{harm gauge}
\eeq
Using condition (\ref{harm gauge}) in Eq.~(\ref{Eq-lin-S}),
the last is cast in the form (see, e.g., \cite{Weinberg})
\beq
\cx h_{\mu\nu}
\,\,=\,-\, 16 \pi G \, S_{\mu\nu}.
\label{Eq-lin-S-harm}
\eeq
For the plane wave propagating along an arbitrary axis, the
components of the metric perturbation (\ref{h_mn}) which are
gauge invariant and can be called physical, are transverse ones,
${\bar h}^{\bot\bot}_{\mu\nu}$. Thus, the gravitational
wave in GR is a propagation of the spin-2 state.

The emission of the gravitational wave in the linear regime
corresponds to the solution in the standard form of retarded
potential,
\beq
h_{\mu\nu} ({\vec x},t)
\,\,=\,\,\frac{4}{M_P^2}\int d^3x'\,\,
\frac{S_{\mu\nu}({\vec x}',
\,t-\left| {\vec x}-{\vec x}'\right|)}
{\left| {\vec x}-{\vec x}'\right|}\,.
\label{gw7}
\eeq
The factor $\, 1/M^2_P = G\,$ in this expression shows that the
emission of the gravitational waves is suppressed by the square
of the Planck mass. And after the wave travels a very long
distance, there is a similar Planck suppression at the moment of its
detection, that explains the difficulty of detecting the gravitational
wave. The remarkable detection by LIGO is explained by the
incredible quantity of energy emitted in the merger of two black
holes or other extremely compact and massive objects, such as
neutron stars.

\subsection{Quantization and the notion of graviton }
\label{IIQG-sec-graviton}

At quantum level, the physical degrees of freedom corresponding
to the state of a free gravitational field on a flat background
correspond to the degrees of freedom of the linear gravitational
wave described above. The corresponding particle with zero mass
and spin-2 is called a graviton.

To derive the spin-2 part of the propagator of the gravitational
perturbation $h_{\mu\nu}$, we can use the tensor part of the
propagator [see, e.g., Eq.~(\ref{Spin2-QG}) below] and setting
$\Phi = 0$. For the sake of simplicity, we omit the coefficient
$\ka^2$ and obtain the spin-2 part of the Euclidean propagator
in the form
\beq
\langle h_{\mu\nu}h_{\al\be}\rangle^{(2)}
\,=\,
G^{(2)}_{\mu\nu , \al\be}(k)
\,=\,
\frac{P^{(2)}_{\mu\nu , \al\be}(k)}{k^2}\,,
\label{prop graviton}
\eeq
where the projector $P^{(2)}_{\mu\nu , \al\be}(k)$ to the spin-2
states will be defined below. This equation describes the propagation
of the massless degrees of freedom associated with the spin-2 states
in GR, which is the graviton.

In other models of QG, the propagator can be more complicated
because  $\Phi$ (and also $\Psi$, because there may be a relevant
dynamics in the scalar sector) are typically nonzero. For instance,
including the fourth-derivative terms, there may be the following
two changes:

i) Instead of the unique massless pole in the propagator
(\ref{prop graviton}), there may be additional massive
poles.

ii) The scalar components of  the metric perturbation (\ref{h_mn})
gain a massive, gauge-independent sector in the propagator.

Adding more derivatives, which means using polynomial or
even nonlocal models, the modifications always concern the
same two points. Namely, there will be (in the polynomial
models) a growing number of poles in the spin-2 sector and
the scalar sector. In contrast, some choices of nonlocal action
may provide that there would not be any massive poles in the
tree-level propagator, in both  spin-2 and  spin-0
sectors.

It is worth noting that the count of degrees of freedom, based on
the simple analysis of the gravitational propagator, was confirmed
by the canonical quantization of the gravitational theory in the
cases of quantum GR and fourth-derivative quantum gravity (see,
e.g., \cite{book}).

\section{Propagator of metric and the Barnes-Rivers
projectors}
\label{secQG1.3}

At this point we note a common aspect of all mentioned models
of QG, namely the ones based on the actions (\ref{EH}),
(\ref{action4der}), (\ref{act Phi}). Since the propagator of
$h_{\mu\nu}$ in flat background is defined by the quadratic in
curvatures (Riemann, Ricci or scalar $R$) terms in the action,
the form factors depending on $\cx$ do not influence the
\textit{tensor structure} of the bilinear form.

The propagator $G$ of the quantum metric, in any QG model,
obeys the equation
\beq
&&
H^{\mu\nu, \al\be}(x)\,
G_{\al\be,\rho\si}(x,y)
\,\,=\,\,\de^4(x-y) \,\de^{\mu\nu}_{\quad,\rho\si}\,,
\label{HG}
\eeq
where
\beq
&&
H^{\mu\nu, \al\be}(x)\,\de^4(x-y)
\,\,=\,\,
\frac{1}{2\sqrt{-g}}
\,\frac{\de^2 S}{\de g_{\mu\nu}(x)\,\de g_{\al\be}(y)}.
\label{biliH-QG}
\eeq
is the bilinear in quantum fields form of the classical action $S$
of a model of quantum gravity. The action in (\ref{biliH-QG})
should include the gauge-fixing term, as otherwise we meet a
degeneracy. For the sake of definiteness, we start from the
gauge-invariant action and denote  the corresponding degenerate
bilinear form coming from the initial action, as
$H_{(0)}^{\mu\nu, \al\be}$, while the non-degenerate
version, after the Faddeev-Popov procedure, will be denoted as
$H^{\mu\nu, \al\be}$.

In this part of the Chapter, all spacetime indices are raised and
lowered with the flat background metric. Then, independently
of  the model and the choice of the gauge fixing, the operator
$H^{\mu\nu, \al\be}(x)$ has  the following tensor structure:
\beq
&&
H_{\mu\nu, \al\be}(x;g)
= a_1  \de_{\mu\nu,\al\be}\square
+ a_2 \eta_{\mu\nu} \eta_{\al\be}\square
+ a_3 \big(\eta_{\mu\nu}\pa_\al \pa_\be
                       + \eta_{\al\be}\pa_\mu \pa_\nu\big)
\nonumber
\\
&&
\qquad
+\, a_4 \big(\eta_{\mu\al}\pa_\be \pa_\nu
+ \eta_{\nu\al}\pa_\be \pa_\mu
+ \eta_{\mu\be}\pa_\al \pa_\nu
+ \eta_{\nu\be}\pa_\al \pa_\mu \big)
\,-\, a_5
\pa_\al \pa_\be \pa_\mu \pa_\nu,
\quad
\label{biliH}
\eeq
where $a_k=a_k(- \square)$ are five model-dependent functions
of the d'Alembert operator. In the higher-derivative cases, all of
these functions are proportional to the linear combinations of
$\Phi(\square)$ and $\Psi(\square)$ in Eq.~(\ref{act Phi}).
In particular, for the fourth-derivative model (\ref{action4der}),
$a_{1,2,3,4}$  are linear functions of $\square$  and $a_5=const$.
In the case of quantum GR, there are the constant functions
$a_{1,..,4}$ and $a_5=0$.
In what follows, we consider the general analysis of the propagator,
which is valid for all types of models.

Making a Fourier transform, we can rewrite the bilinear form in
the momentum representation,
\beq
&&
H_{\mu\nu, \al\be}(k;\eta)
\,=\,
- \big[a_1(k^2)  \de_{\mu\nu,\al\be}k^2
+ a_2(k^2) \eta_{\mu\nu} \eta_{\al\be} k^2
\nonumber
\\
&&
\quad
+\,
a_3(k^2) \big(\eta_{\mu\nu}k_\al k_\be
+ \eta_{\al\be}k_\mu k_\nu\big)
\mbox{\,\,}
\label{biliH-k}
\\
&&
\quad +\,
a_4(k^2) \big(\eta_{\mu\al}k_\be k_\nu
+ \eta_{\nu\al}k_\be k_\mu
+ \eta_{\mu\be}k_\al k_\nu
+ \eta_{\nu\be}k_\al k_\mu \big)
+ a_5(k^2)
k_\al k_\be k_\mu k_\nu \big],
\nn
\eeq
where $k^2=k_\mu k^\mu$ is the square of the four-dimensional
momentum and $\de_{\mu\nu,\al\be}$ is similar to (\ref{delta_abmn}),
but this time it is constructed from the flat metric $\eta_{\mu\nu}$.

It is useful to present (\ref{biliH-k}) in a slightly different form,
providing more generality by using the $n$-dimensional versions
of the  formulas,
\beq
&&
{\hat H}
\,=\,
s_1\,{\hat T}_1 \,+\, s_2\,{\hat T}_2 \,+\, s_3\,{\hat T}_3
\,+\, s_4\,{\hat T}_4 \,+\, s_5\,{\hat T}_5,
\label{biliH-k-mod}
\eeq
where \ \
${\hat T}_n \,=\,T^{(n)}_{\mu\nu,\al\be}$ \ \ and
\beq
&&
{\hat T}_1 \,=\, \de_{\mu\nu,\al\be},
\qquad
{\hat T}_2 \,=\, \eta_{\mu\nu} \eta_{\al\be},
\qquad
{\hat T}_3 \,=\, \frac{1}{k^2}
\big( \eta_{\mu\nu} k_\al k_\be + \eta_{\al\be} k_\mu k_\nu \big),
\label{biliH-T}
\\
&&
{\hat T}_4 \,=\, \frac{1}{4k^2}
\big(\eta_{\mu\al}k_\be k_\nu
+ \eta_{\nu\al}k_\be k_\mu
+ \eta_{\mu\be}k_\al k_\nu
+ \eta_{\nu\be}k_\al k_\mu \big)\,,
\quad
{\hat T}_5\,=\, \frac{1}{k^4}\, k_\al k_\be k_\mu k_\nu .
\nonumber
\eeq
The coefficients depend on momentum, $s_l=s_l(k^2)$, and these
dependencies may be nontrivial, e.g., in the polynomial or nonlocal
models. However, the tensor structure of the expressions
(\ref{biliH-k-mod}) is the same for all QG models, i.e., for quantum
GR, or for a higher-derivative polynomial, or nonlocal models
(\ref{act Phi}).

To invert the operator (\ref{biliH-k}) and take care about its
possible degeneracy, consider the operators called Barnes-Rivers
projectors \cite{Barnes,Rivers}. The starting point is to formulate
the projectors to the transverse and longitudinal subspaces of the
vector space. In the momentum representation we have
\beq
\om_{\mu\nu} = \frac{k_\mu k_\nu}{k^2}
,\qquad
\th_{\mu\nu} = \eta_{\mu\nu} - \om_{\mu\nu},
\label{projvect}
\eeq
with the standard properties
\beq
\om_{\mu\nu}\, \om^\nu_{\,\,\,\,\la} = \om_{\mu\la}
,\qquad
\th_{\mu\nu}\, \th^\nu_{\,\,\,\,\la} = \th_{\mu\la}
,\qquad
\om_{\mu\nu}\, \th^\nu_{\,\,\,\,\la} = 0.
\label{projvect-props}
\eeq
Then, the projectors to the spin-2, spin-1 and spin-0 states in the
symmetric tensors space are written in the form
\beq
{\hat P}^{(2)}
&=&
P^{(2)}_{\mu\nu\,,\,\al\be}
\,=\,
\frac12(\theta_{\mu\al}\theta_{\nu\be}
+ \theta_{\mu\be}\theta_{\nu\al})
- \frac{1}{n-1}\,\theta_{\mu\nu}\theta_{\al\be}\,,
\nonumber
\\
{ \hat P}^{(1)}
&=&
P^{(1)}_{\mu\nu\,,\,\al\be}
\,=\,
\frac12\,(\theta_{\mu\al}\om_{\nu\be} + \theta_{\nu\al}\om_{\mu\be}
+ \theta_{\mu\be}\om_{\nu\al} + \theta_{\nu\be}\om_{\mu\al})\,,
\nonumber
\\
{\hat P}^{(0-s)}
&=&
P^{{(0-s)}}_{\mu\nu\,,\,\al\be}
\,=\,
\frac{1}{n-1}\,\theta_{\mu\nu}\theta_{\al\be},
\quad
{\hat P}^{(0-w)}
\,=\,
P^{{(0-w)}}_{\mu\nu\,,\,\al\be}
\,=\,
\om_{\mu\nu}\om_{\al\be}.
\mbox{\qquad}
\label{proj-tensor}
\eeq
On top of these, to have the closed algebra of projectors in the
scalar sector, one needs the additional transfer operators
\beq
{\hat P}^{{(ws)}}
= P^{(ws)}_{\mu\nu\,,\,\al\be}
= \frac{1}{\sqrt{n-1}}\,\theta_{\mu\nu}\om_{\al\be}
,\quad
{\hat P}^{{(sw)}}
= P^{(sw)}_{\mu\nu\,,\,\al\be}
= \frac{1}{\sqrt{n-1}}\,\om_{\mu\nu}\theta_{\al\be}.
\mbox{\qquad}
\label{projBR}
\eeq

The algebra for the vector and tensor projectors is simple,
\beq
{\hat P}^{(2)} \,{\hat P}^{(i)} \,=\, {\hat P}^{(2)}\de_{i2}
\qquad
\mbox{and}
\qquad
{\hat P}^{(1)} \,{\hat P}^{(i)} \,=\, {\hat P}^{(1)}\de_{i1}\,,
\label{ap1}
\eeq
where $i=(2,\,\,1,\,\,0-w,\,\,0-s,\,\,sw,\,\,ws)$.
In the scalar sector, one has to construct the matrix projector
operator
\beq
{\hat P_0} \,\,=\,\, \frac12\,
\left\|
\begin{array}{cc}
P^{{(0-s)}}   &  P^{{(sw)}}  \\
P^{{(ws)}}  &  P^{{(0-w)}}
\end{array}
\right\|
\label{ap2}
\eeq
satisfying the relation ${\hat P_0}^2\,=\, {\hat P_0}$. To end
this part, the last two terms, which represent the scalar sector of
(\ref{h_mn}), can be written, in momentum representation, as
\beq
h_{\mu\nu}^{scalar}
\,\,=\,\,
\frac14\,h \,\th_{\mu\nu}
\,+\,
\Big(\frac14\, h - k^2 \ep\,\Big)\om_{\mu\nu},
\label{h-mn-scal}
\eeq
such that acting by each of the two projectors $P^{{(0-s)}}$
and $P^{{(0-w)}}$, one of these terms remains invariant and another
vanish.

Now we are in a position to find the propagator.
Solving Eq.~(\ref{HG}) requires the inversion of the expression
\beq
{\hat H}
\,=\,
b_2{\hat P}^{(2)} + b_1{\hat P}^{(1)}
+ b_{os} P^{{(0-s)}}+ b_{ow}{\hat P}^{(0-w)}
+ b_{sw}\big[  P^{{(ws)}}+ {\hat P}^{(sw)} \big],
\mbox{\qquad}
\mbox{\,\,}
\label{bigB}
\eeq
that means one has to find such an operator
\beq
{\hat G}
\,=\,
c_2{\hat P}^{(2)} + c_1{\hat P}^{(1)}
+ c_{os} P^{{(0-s)}}+ c_{ow}{\hat P}^{(0-w)}
+ c_{sw}\big[  P^{{(ws)}}+ {\hat P}^{(sw)} \big],
\mbox{\qquad}
\mbox{\,\,}
\label{bigC}
\eeq
that the product with ${\hat B}$ is unity,
\beq
{\hat H}\,{\hat G}
\,=\,{\hat 1}\,=\,
{\hat P}^{(2)} + {\hat P}^{(1)}+ P^{{(0-s)}}+{\hat P}^{(0-w)}.
\label{1m}
\eeq
Using the aforementioned algebra of projectors, we get
the solution to this problem,
\beq
&&
c_2 = \frac{1}{b_2}, \quad
c_1 = \frac{1}{b_1}, \quad
c_{os} = -\,\frac{b_{ow}}{\De}, \quad
c_{ow} = -\,\frac{b_{os}}{\De}, \quad
c_{sw} = \frac{b_{sw}}{\De},
\label{B2C}
\eeq
where $\quad\De = b^2_{sw} - b_{os}b_{ow}$.

It is clear that the action of a generally covariant theory before
adding the gauge fixing term has either $b_1=0$ or $\De=0$.
From this perspective, the purpose of the Faddeev-Popov
procedure is to remove this degeneracy.

It remains to present the projectors (\ref{proj-tensor}) and
the transfer operators (\ref{projBR}) as the linear combinations
of the expressions (\ref{biliH-T}), and \textit{v.v.}. The first set
of formulas is
\beq
&&
{\hat P}^{(2)}
\,\,=\,\, {\hat T}_1 \,-\, \frac{1}{n-1}\,{\hat T}_2
\,+\, \frac{1}{n-1}\,{\hat T}_3
\,-\,2{\hat T}_4
\,+\, \frac{n-2}{n-1}\,{\hat T}_5,
\qquad\quad
{\hat P}^{(1)}
\,\,=\,\,
\,2\big({\hat T}_4 \,-\, {\hat T}_5\big),
\nn
\\
&&
{\hat P}^{(0-w)}
\,\,=\,\, {\hat T}_5 ,
\qquad
{\hat P}^{(0-s)}
\,\,=\,\,
\frac{1}{n-1}\,
\big( {\hat T}_2\,-\, {\hat T}_3 \,+\, {\hat T}_5\big),
\nn
\\
&&
P^{{(ws)}} + P^{{(sw)}}
\,\,=\,\,
\frac{1}{\sqrt{n-1}}\,
\big( {\hat T}_3 - 2 {\hat T}_5\big).
\label{PT}
\eeq
Finally, by inverting these relations, one can express the matrices
in (\ref{biliH-T}) as
\beq
&&
{\hat T}_1 \,=\, {\hat P}^{(2)} \,+\, {\hat P}^{(1)}
\,+\,  P^{{(0-s)}}\,+\, {\hat P}^{(0-w)},
\nonumber
\\
&&
{\hat T}_2 \,=\, (n-1)P^{{(0-s)}}
\,+\, \sqrt{n-1}\big[  P^{{(ws)}}\,+\, {\hat P}^{(sw)} \big]
\,+\, {\hat P}^{(0-w)},
\nonumber
\\
&&
{\hat T}_3
\,=\, \sqrt{n-1}\big[  P^{{(ws)}}\,+\, {\hat P}^{(sw)} \big]
\,+\, 2{\hat P}^{(0-w)},
\nonumber
\\
&&
{\hat T}_4 \,=\, \frac{1}{2}\,{\hat P}^{(1)} \,+\, {\hat P}^{(0-w)},
\qquad
{\hat T}_5 \,=\, {\hat P}^{(0-w)}.
\label{TP}
\eeq

Correspondingly, the transformations between the coefficients of
(\ref{biliH-k-mod}) and (\ref{bigB}) are given by the inverse
relations to the ones of ``basic vectors'', i.e.,
\beq
&&
b_2 \,=\, s_1 \,+\, s_2 \,+\, s_3 \,+\, s_4,
\qquad
b_1\,=\, (n-1)s_3 \,+\, \sqrt{n-1}s_5 \,+\, s_4,
\nonumber
\\
&&
b_{0s}\,=\, \sqrt{n-1}s_5 \,+\, 2s_4,
\qquad
b_{0w}\,=\,  \frac{1}{2}\,s_2 \,+\, s_4,
\qquad
b_{sw}\,=\, s_4.
\label{sb}
\eeq
and
\beq
&&
s_1 \,=\, b_2 \,-\, \frac{1}{n-1}\,b_1
\,+\, \frac{1}{n-1}\,b_{0s}
\,-\, 2 b_{0w} \,+\, \frac{n-2}{n-1}\,b_{sw},
\qquad
s_2 \,=\, \big(b_{0w} - b_{sw}\big),
\nonumber
\\
&&
s_3 \,=\, \frac{1}{n-1}\big(b_1 - b_{0s} + b_{sw}\big),
\quad
s_4 \,=\, b_{sw},
\quad
s_5 \,=\, \frac{1}{\sqrt{n-1}}\big(b_{0s} - 2 b_{sw}\big).
\label{bs}
\eeq

The solution to Eq.~(\ref{HG}) consists of casting the bilinear form
in the standard form (\ref{biliH-T}), using the relations between
${\hat T}_l$ and projectors (\ref{TP}), inverting the result using
(\ref{B2C}) and, finally, using the inverse relations (\ref{PT}).
In principle, this procedure works for the bilinear form of the total
action (\ref{biliH}) for an arbitrary model of quantum gravity if the
chosen gauge-fixing term makes the bilinear form non-degenerate.
Usually, the
Faddeev-Popov procedure is sufficient in this respect, but if the
original theory had an extra symmetry (e.g., the conformal one), one
needs to apply an additional gauge fixing, e.g., setting
$h=h^\mu_{\,\,\mu}=0$ \cite{frts82}.

To put  the described procedure in practise, one has to use
definitions (\ref{C2n}), with $\cx^n \to \Phi$, and (\ref{GBterm}),
and insert the form factors $\Phi$, $\Psi$ and $\Om$ into expansions
(\ref{Rim2-expand}),  (\ref{Ricci2-expand}), (\ref{R2-expand}), and
(\ref{EH-expand}). The bilinear form of the general gauge-fixing term
(\ref{Sgf-QG}) with the weight (\ref{weight-nonloc}), can be
easily calculated,
\beq
H^{(GF)} _{\mu\nu, \al\be}(k;\,\eta)
\,=\,{\hat H}_{GF}
&=&
W(-k^2)\,k^4
\bigg\{\be^2(\ga-1){\hat T}_2
\,+\,\be(1-\ga){\hat T}_3
\,-\,\frac14{\hat T}_4
\,+\,\ga\, {\hat T}_5\bigg\}.
\nonumber
\\
&&
\label{biliH-gf}
\eeq

Summing up all the terms, including the bilinear form of the original
action $H_{(0)\,\mu\nu, \al\be}(k;\,\eta)$, contribution of
(\ref{GBterm}) and of the gauge-fixing term, we arrive at the
expression (\ref{biliH-k-mod})
\beq
&&
{\hat H}
\,=\,
{\hat H}_{(0)}
\,+\,
{\hat H}_{GB}
\,+\,
{\hat H}_{GF},
\label{biliH-tot}
\eeq
with the coefficients
\beq
&&
s_1 \,=\, \frac12\, \Phi k^4
+ \frac{1}{2\ka^2}\, k^2
+ \frac{\La}{\ka^2}\,,
\nn
\\
&&
s_2 \,=\,
\Big(\Psi - \frac16\, \Phi\Big) k^4
- \frac{1}{2\ka^2}\, k^2
- \frac{\La}{2\ka^2}
+ \be^2 (\ga - 1) Wk^4,
\nn
\\
&&
s_3 \,=\,
\Big(\frac16\,\Phi -\Psi\Big) k^4
+ \frac{1}{2\ka^2}\, k^2
+ \be (1-\ga) Wk^4,
\nn
\\
&&
s_4 \,=\, -\,\Phi k^4
- \frac{1}{\ka^2}\, k^2
- Wk^4,
\nn
\\
&&
s_5 \,=\,
\Big(\frac13\,\Phi + \Psi \Big) k^4
+ \ga Wk^4.
\label{ss}
\eeq
The remarkable detail is that there is no $\Om(-k^2)$ in these
expressions. This is certainly an expected result because this
function comes from the ``generalized'' topological invariant
(\ref{GBterm}), but we observe that this feature holds for any
$\Om(-k^2)$, not only for a constant, when this term in the
action is really topological.

Another expected characteristic of expressions (\ref{ss}) is that all
the coefficients except $s_1$ depend on the gauge fixing parameters.
Since (\ref{biliH-tot}) has three linearly
independent coefficients $\frac{1}{4\al}$,
$\frac{\be\ga}{\al}$ and $\frac{1-\ga}{\al}$, by using the choice
of the three parameters $\al$, $\be$ and $\ga$, one can eliminate
terms with the coefficients $a_{3,4,5}$ in the bilinear form
(\ref{biliH-tot}). As a result, in any QG model, one can provide
the minimal form of the total bilinear operator,
\beq
&&
H^{total\,,\,minimal}_{\mu\nu, \al\be}(k;\,\eta)
\,=\,
- \,\Big[
a_1(k^2)  \de_{\mu\nu,\al\be}
\,+\, a'_2(k^2) \eta_{\mu\nu} \eta_{\al\be}\Big] k^2,
\label{biliH-k-tot}
\eeq
where $a'_2$ differs from $a_2$ in Eq.~(\ref{biliH-k})
because of the contribution of the gauge-fixing term. This property
of the bilinear form holds also for an arbitrary background metric
and is important for the one- or higher-loop calculations in QG.

The main advantage of the minimal form (\ref{biliH-k-tot}) compared
to the general one, is that the minimal operators are directly suited
for the use of the heat-kernel technique. Indeed, it is possible to
work with the nonminimal operators in quantum gravity, e.g.,
using the generalized Schwinger-DeWitt technique \cite{bavi85},
but it is always simpler to work with the minimal bilinear forms.

In some QG models, the choice of parameters is more restricted.
E.g., in quantum GR, we meet only two gauge-fixing parameters,
i.e., $\al$ and $\be$, but there are only four nonminimal terms
because of $a_5=0$. As a result, one can choose $\al$ and $\be$
to achieve the minimal form (\ref{biliH-k-tot}). One technical
observation is that, if the initial action includes, simultaneously,
higher derivative ($\Phi$- and $\Psi$-terms) and the Einstein-Hilbert
action, one can provide minimality only in the higher order terms
(typically, in all of them, greatly simplifying calculations in the
superrenormalizable models), but not in the second derivative
sector of the operator.

Let us now analyse the situation from another perspective.
Expressions  (\ref{ss}) remain the same in any dimension $n$.
Thus, we can use (\ref{sb}) to transform the operator ${\hat H}$
into the form (\ref{bigB}). The result of this transformation is
\beq
&&
b_2 \,=\, \frac12\, \Phi k^4
+ \frac{1}{2\ka^2}\, k^2
+ \frac{\La}{\ka^2}\,,
\nn
\\
&&
b_1 \,=\, - \frac{1}{2}\,Wk^4 + \frac{\La}{\ka^2},
\nn
\\
&&
b_{0s} \,=\,
- \frac{n-4}{6}\,\Phi\,k^4
 + (n-1) \big[\Psi +  \be^2 (\ga -1) W \big]k^4
   \,-\,
\frac{n-2}{2 \ka^2}\,k^2
   \,-\,
\frac{ (n-3)\La}{2\ka^2}\,,
\nn
\\
&&
b_{0w} \,=\,
(1 - \be)^2 (\ga - 1)W k^4
\,+\, \frac{k^2}{\ka^2}
\,+\, \frac{\La}{2\ka^2}\,,
\nn
\\
&&
b_{sw} \,=\,\sqrt{n-1}\,
\Big[\be (\be-1) (\ga -1) k^4 W
- \frac{\La}{2\ka^2}
\Big]\,.
\label{bb}
\eeq
Once again, only the spin-2 coefficient is gauge-fixing independent.

The inversion formulas (\ref{B2C}) are trivial in the spin-2 and
spin-1 sectors. In the tensor sector, we get, for the $\La = 0$ case,
independent on the dimension $\,n\,$ and on the gauge fixing,
\beq
&&
c_2(\La = 0)
\, \, \, = \, \, \,
\frac{2\ka^2}{k^2\big(1 + \ka^2 k^2\Phi\big)}.
\label{Spin2-QG}
\eeq
The vector part depends on the gauge fixing and has no direct
physical interpretation. Thus, we concentrate on the results in
the scalar sector.
The reader can easily obtain the complete formulas, but since these
formulas are cumbersome, we shall present only the qualitative
results and the most interesting expression. In the cases of
$\La \neq 0$ or $n\neq 4$, all the
scalar coefficients are gauge-fixing dependent. However, in case
 $\La = 0$ and $n = 4$, one important coefficient is
 invariant\footnote{I am grateful to Dr. Leslaw Rachwa\l \ for
indicating to me this feature.},
\beq
&&
c_{os}(\La = 0,\,n = 4)
\, \, \, = \, \, \,
-\,\frac{\ka^2}{k^2\big(1 - 3\ka^2 \Psi k^2 \big)}.
\label{B2C-QG}
\eeq
It is easy to see from these formulas, that the propagator
(\ref{Spin2-QG}) of the spin-2 mode depends only on the function
$\Phi$, while the spin-0 propagator  (\ref{B2C-QG}) depends only
on the function $\Psi$. Let us note that this output was anticipated
already at the level of bilinear expansions of the classical action.
Furthermore, both spin-2 and spin-0 propagators are gauge-fixing
independent and contribute to the tree-level $S$-matrix of
gravitational perturbations. It is interesting that these features,
which are looking quite special, hold independent on the form
of the form factors $\Phi(\cx)$ and $\Psi(\cx)$ in the action
(\ref{act Phi}). The same concerns the irrelevance of the third form
factor  $\Om$ for the propagator. Of course, this property is not
valid for the vertices, if  $\Om(x)$ is not a constant function.

A peculiar situation occurs in the usual second-derivative gravity.
Even if one starts from
the pure GR, the result is the same as if setting $\,\Psi \to 0\,$ in
(\ref{B2C-QG}), i.e.,\footnote{The factor of $\ka^2$ in this
formula appears because we used the expansion (\ref{bfm}) with
a flat background $g_{\mu\nu}=\eta_{\mu\nu}$. If using the
expansion (\ref{qg-flat}), there is no such coefficient.}
\beq
c_{os}(GR\mbox{  \ with \ }\La = 0)
\, \, \, = \, \, \, -\,\frac{\ka^2}{k^2}.
\label{B2-GR-QG}
\eeq
This formula is in apparent contradiction with what we saw in the
analysis of the gravitational waves in GR, where the unique sort
of the propagating degrees of freedom are the tensor (transverse and
traceless) modes. The explanation of this apparent discrepancy is
that the smooth $\,\Psi \to 0\,$ limit in (\ref{B2-GR-QG})
corresponds to the theory after the Faddeev-Popov procedure,
which extends the space of the propagating modes and make the
whole propagator non-degenerate. On the contrary, the result for
the gravitational waves is based on another procedure, i.e.,
removing all degrees of freedom by using gauge transformation
and its remnant (see, e.g., \cite{Weinberg}).

From the QG perspective, the smooth limit (\ref{B2-GR-QG}) of the
general expression (\ref{B2C-QG}) for the scalar sector is important,
as it provides universal IR limit of the propagator (for $n=4$ only!)
in any QG model (\ref{act Phi}) at the tree level, i.e., in both relevant
sectors of the propagator.

\section{Gauge-invariant renormalization in quantum gravity}
\label{IIQG-sec1.4}

In the next Chapter of this Section there is a detailed proof of the
two main statements concerning the gauge-invariant
renormalization in quantum gravity \cite{Lavrov-renQG}
(see also pioneer work \cite{Stelle77}, \cite{VorTyu-QG}
and more recent and \cite{BarvSib-QG,Ohta}) .

Both theorems were already mentioned in the Introduction. However,
since these two statements are relevant for the rest of this Chapter,
let us formulate them here in more detail.
\vskip 1mm

1. The renormalization preserves the diffeomorphism invariance
(general covariance) of the model of QG in four
spacetime dimensions (i.e., in $n=4$), if the initial classical
theory possesses this symmetry. This means, one can remove
the divergences in all loop orders by adding covariant counterterms.
This statement applies literally only to the divergences that take
place in the framework of the background field method, which is
especially designed to avoid non-covariant counterterms. Let us
stress that the background field method is not necessary for the
gauge-invariant renormalization in QG. However,
using the non-covariant parametrization of the metric, such as
(\ref{qg-flat}), one has to go through a relatively complicated
procedure or additional renormalization transformations, as
described in \cite{Stelle77}. The final output is always the same
in the sense of the same essential covariant counterterms.
For this reason, in what follows we shall assume (\ref{qg-flat})
when evaluating the power counting in different models of QG.
However, we shall switch to the more general parametrization with
the general background metric (\ref{bfm}) when making the
practical calculations.
\vskip 1mm

2. The dependence on the choice of the gauge condition
(\ref{Sgf-QG}),  (e.g., on the parameters $\be$, $\ga$ and
the function $W(\cx)$ in the weight operator (\ref{weight-nonloc})
is proportional to the effective equations of motion. The same
concerns the dependence of the parametrization of the quantum
metric. In particular, both ambiguities vanish on the classical
mass shell for the one-loop divergences of the effective action.
In what follows (see more details in \cite{OUP}) we shall
demonstrate how this feature works in the practical calculations
in QG.
\vskip 1mm

On top of that, there is the third statement, representing the
general feature of QFT and valid, in particular, for QG. The
counterterms required to remove UV divergences, in all loop
orders, are local functionals of the fields. The mathematically
rigid proof of this statement (usually called Weinberg's theorem
\cite{Weinberg1960}) is complicated and can be found, e.g., in
\cite{Collins}. In what follows, we shall apply these three
statement to describe the renormalization in QG.

\section{Power counting, and classification of
quantum gravity models}
\label{secQG1.4}

To estimate the power counting for the Feynman diagrams in QG is
somehow simpler than in the quantum theories of other fields. The
reason is that
the metric is a dimensionless field. For this reason, the dimensions
of the counterterm that emerge for diagram $G$, with $L$ loops, is
defined only by the number of derivatives of the background metric
or, in case of a non-covariant parametrization such as
(\ref{qg-flat}), by  the number of derivatives of $h_{\mu\nu}$. The
power counting of a diagram is essentially equivalent to the count of
dimensions and, therefore, it does not depend on the number of
external lines of $h_{\mu\nu}$. Let us use the last version of the
expansion, but with $\ka \to 1$ for generality, as this enables us to
include the higher derivative theories into consideration.

The superficial degree of divergences will be denoted $\omega(G)$
(sometimes it is also called the index of divergence) and $d(G)$ is
the number of partial derivatives of the external lines of the field
$h_{\mu\nu}$ in the diagram. Taking into account the powers of
momenta in all elements of the diagrams (see, e.g., \cite{OUP} for
a detailed general treatment), the general expression is
\beq
\om(G)+d(G)
\,=\, \sum\limits_{l_{int}}(4 - r_I)
\,-\,4V \,+\,4\, + \,\sum\limits_{V}K_V,
\label{omG-qg}
\eeq
where the first sum is over all $I$ internal lines of the diagram,
$r_I$ is the inverse power of momentum in the propagator of an
internal line and $V$ is the number of vertices. The last sum is
taken over all the vertices, where $K_V$ is the power of momenta,
(or number of derivatives, in the coordinate representation) of all
the lines coming to the given vertex.

It is easy to see that formula (\ref{omG-qg}) is insufficient to
evaluate the renormalizability of the given QFT or QG model.
In addition to this formula, there is the simple topological
relation
\beq
L=I-V+1
\label{topa}
\eeq
valid for all the relevant diagrams.

Before going on to consider concrete models of QG, let us make the
following observation. The diagrams in quantum gravity, which we
intend to analyze, have external lines of the field $h_{\mu\nu}$
only, but there are internal lines of both  $h_{\mu\nu}$ and the
Faddeev-Popov ghosts. However, with the modified definitions of
the ghost actions (\ref{Mghost-mod}), the values of $r_I$ are the
same for both kinds of quantum fields. E.g., in quantum GR, in
both cases we have $r_I=2$, in fourth-derivative gravity in both cases
$r_I=4$, in the polynomial models  $r_I=2N+4$. Finally, in the
nonlocal QG models (\ref{act Phi}), both  $r_I$ and $K_V$ are
infinite for both metric and ghosts. Then the use of the combination
of (\ref{omG-qg}) and (\ref{topa}) is not possible. However, we shall
see how to deal with this special case using the topological relation
(\ref{topa}) alone. For a while, we assume that in all models of
interest, $r_I$ are identical for the quantum metric and the ghosts.

\subsection{Power counting in quantum gravity based on GR}

As the first step, consider power counting in quantum GR, where
with $r_I=2$ for all internal lines. The vertices coming from the
Einstein-Hilbert term have $K_{EH}=2$. If we include the cosmological
constant term, there are also vertices $K_\La=0$. However, looking
only for strongest divergences, at first we consider only the diagrams
with $K_V=2$ vertices. Then (\ref{omG-qg}), together with the
topological relation (\ref{topa}),  yields
\beq
\om(G)+d(G)
\,=\, 2I -4V + 4 + 2V
\,=\, 2I - 2V + 4
\,=\, 2 + 2L.
\mbox{\qquad}
\label{omG-GR}
\eeq
The last result clearly shows that the QG based on GR
is non-renormalizable. At one-loop $L=1$ and the logarithmic
divergences with $\,\om(G)=0\,$ have $\,d(G)=4$. Taking into
account the diffeomorphism invariance, this indicates towards the
counterterms repeating the covariant structures included in the
fourth-derivative action
(\ref{action4der}). Indeed, at the one-loop order, there are
counterterms of the Einstein-Hilbert form $\sim \int \sqrt{-g}R$ with
$d(G)=2$ but with quadratic divergences only, since $\om(G)=2$.

The logarithmic divergences of this type are also possible, but only
if we introduce cosmological constant term. If there is one vertex
with $K_\La=0$, the diagram produces the logarithmic divergence
with two derivatives.
Assuming the covariance, this means an Einstein-Hilbert type
counterterm. With two such vertices, we meet a logarithmic
divergence without derivatives, i.e., with $d(G)=0$. In one of the
next sections, we  confirm these
conclusions by direct calculations and also analyze the gauge-fixing
and parametrization dependence of the one-loop counterterms.

One can rewrite the one-loop divergences using the relations
\beq
C^2 = E_4 + 2W \,\,\,
\big(\mbox{where}
\quad
W = R_{\mu\nu}^2 - \textstyle{\frac13}R^2 \big), \quad
E_4,\quad
R^2, \quad
\square R.
\label{1-loop 4D}
\eeq
We know that  $E_4$ and $\square R$ are surface terms, which
do not affect the dynamics of the theory, and the other two terms
vanish on the classical equations of motion, when $R_{\mu\nu}=0$.
Thus, the one-loop $S$-matrix in the pure quantum GR (pure means
without matter contents) is finite. In the presence of matter this
feature doesn't hold \cite{hove,dene}, but let us concern only
pure QG.

In the two-loop order  $L=2$. According to Eq.~(\ref{omG-GR}),
the logarithmic divergences without $K_\La$ vertices have
dimension six.  A complete list of the corresponding terms has
been elaborated in the works on the conformal anomaly in six
spacetime dimensions \cite{Bonora}. This list includes
\begin{align}
&\Si_1 = R_{\mu \nu} R^{\mu \al}R_{\al}^{\nu}
&& \Si_2 =  (\na_{\la}{R}_{\mu \nu \al \be})^2
&& \Si_3 = {R}_{\mu \al \nu \be}\na^{\mu}\na^{\nu} {R}^{\al \be}
\nonumber
\\
&
\Si_4 = {R}_{\mu \nu} {R}^{\mu \la \al \be} {R}^{\nu}\,_{\la \al \be}
&&
\Si_5 = R^{\mu \nu}\,_{\al \be}
R^{\al \be}\,_{\la \ta} R^{\la \ta}\,_{\mu \nu}
&&
\Si_6 =  R^{\mu}\,_{\al}\,^{\nu}\,_{\be}
           R^{\al}\,_{\la}\,^{\be}\,_{\ta}
           R^{\la}\,_{\mu}\,^{\ta}\,_{\nu}
 \nonumber
 \\
 &\Si_7 = (\na_{\la}{R}_{\mu \nu})^2
&& \Si_8 =  {R}_{\mu \nu}{\cx}{R}^{\mu \nu}
&& \Si_9 =  (\na_{\mu}{R})^2
\nonumber
\\
&\Si_{10} = {R}{\cx}{R}
&& \Si_{11} =  (\na_{\al}{R}_{\mu \nu})\na^{\mu}{R}^{\nu \al}
&& \Si_{12} =  R^{\mu \nu} \na_{\mu}\na_{\nu} R
\nonumber
\\
&\Si_{13} =  R_{\mu \nu} R_{\al \be} {R}^{\mu \al \nu \be}
\label{base_A}
\end{align}
as well as the set of surface terms,
\beq
&&
\Xi_1 = {\cx}^2 R
\quad\qquad
\Xi_2 = \Box R^2_{\mu \nu \al \be}
\qquad
\Xi_3 = \Box R^2_{\mu \nu}
\qquad\quad
\Xi_4 = \Box R^2
\nonumber
\\
&&
\Xi_5 = \na_\mu \na_\nu \big(R^{\mu}\,_{\la \al \be} R^{\nu \la \al \be}\big)
\qquad
\Xi_6= \na_\mu \na_\nu \big(R_{\al \be} R^{\mu \al \nu \be}\big)
\nonumber
\\
&&
\Xi_7 = \na_\mu \na_\nu \big(R_{\al}^{\mu} R^{\nu \al}\big)
\qquad
\qquad\quad
\quad\,\,
\Xi_8 =  \na_\mu \na_\nu \big(R R^{\mu \nu} \big),
\label{oito}
\eeq
satisfying the identity
\beq
\label{Xi_comb}
\Xi_2 -4\Xi_3 + \Xi_4 -4\Xi_5+ 8\Xi_6 + 8\Xi_7 -4\Xi_8 \,=\,0.
\eeq
All these structures can show up in the two-loop divergences, but
only two of these terms, namely, $\Si_5$ and  $\Si_6$ are critically
important because they do not vanish on shell. The two-loop
calculation were done in \cite{gosa} and confirmed in \cite{ven}
by using another calculational approach. The results confirmed the
non-zero coefficient of $ \Si_5$. The conclusion is that there are no
miracles and the theory of QG based on GR is non-renormalizable.

Within the standard perturbative approach, the non-renormalizability
means the theory has no predictive power. With every new order of
the loop expansion, there are new types of local covariant
divergences, with the growing number of derivatives of the
metric. And every time when a new type of counterterm is introduced,
it is necessary to fix the renormalization condition. Each of these
conditions requires making a measurement and using its result to
fix the value of the corresponding parameter. In quantum GR, this
sequence of operations is formally infinite. Thus, before making
a single prediction, it is necessary to use an infinite amount of
experimental data.

What are the possible ways out of this situation? The main
options are as follows.

1. \  Change standard perturbative approach to something else.
The reader can consult other Sections of our Handbook, to see
how the problem is solved in the framework of non-perturbative
approaches, superstring theory, etc.  Let us say that there are
many interesting options, but their consistency and the relation
to the QG program are not completely clear, in all cases.

2. \
Restrict the area of application of QG to the low-energy domain.
The reader can read about this possibility in the Section about
effective QG. The main problem with this approach is that the
QG is initially supposed to be a concept describing extreme
high-energy Physics, with the typical energy scale of the Planck
order of magnitude. It is certainly important to know what
remains from the QG effects in the IR, but this does not reduce
the importance of formulating QG that would be applicable at
high energies.

3. \  Change the theory, i.e., start from the model different from
GR, to construct QG. This option is the mainstream direction in
perturbative QG. It is important that the problems we meet in
this way, such as the problem of nonphysical ghosts coming
from higher derivative terms, persist in many non-traditional
approaches which we mentioned in the first point (see, e.g.,
the discussion in \cite{ABSh2}).

\subsection{Power counting in fourth-derivative gravity models}

The next example is the power counting in the fourth-derivative
quantum gravity (\ref{action4der}). We assume the Faddeev-Popov
procedure with the second-order weight operator (\ref{weight-4der})
and the modified action of  ghosts (\ref{Mghost-mod}). In this case,
for all modes of the gravitational perturbation $h_{\mu\nu}$ and
ghosts, we have $r_l=4$, while the vertices $K_{V}$ include $K_{4d}$,
$\,K_{EH}$, and $\,K_\La$.

Let us denote $n_{4d}$ the number of vertices with fourth
power of momenta, $n_{EH}$ the one with two, and $n_\La$
-- with zero power of momenta. Then
\beq
n_{4d}+n_{EH}+n_\La=V
\quad
\mbox{and}
\quad
n_{4d}K_{4d} + n_{EH} K_{EH} + n_\La K_\La=\sum_VK_V.
\qquad
\label{nK-4der}
\eeq
The general expression (\ref{omG-qg}), together with the topological
relation (\ref{topa}), give the following result:
\beq
\om(G)+d(G) \,=\, 4 - 2n_{EH} - 4n_\La\,.
\label{omG-4der}
\eeq
As a starting point, consider the diagrams with the strongest
divergences, where all of the vertices are of the $K_{4d}$ type,
i.e., $V=n_{4d}$. In this case, (\ref{omG-4der}) means that the
logarithmic divergences have $d(G)=4$. Taking into account locality
and covariance arguments, the possible counterterms are of the
$C^2$, $R^2$, $E_4$ and $\square R$ types. This means, in all
loop orders the divergences have the same form as the
fourth-derivative terms in the classical action (\ref{action4der}).
Then, for $n_{4d}=V-1$ and $n_{EH}=1$, we obtain  $d(G)=2$,
corresponding to the counterterm linear in $R$.
Finally, for $n_{EH}=2$ and $V-2=n_{4d}$, or for $n_\La=1$ and
$V-1=n_{4d}$, there is a counterterm with $d(G)=0$, which is the
cosmological constant. Thus, the theory under consideration
is multiplicatively renormalizable. In the next sections, we shall see
that this does not mean that the theory is completely consistent,
as there is a massive nonphysical ghost in the spectrum and the
subsequent problems with quantum  unitarity and even with the
stability of classical solutions.

For the particular case of the general model (\ref{action4der})
without dimensional parameters, the classical action has a global
conformal symmetry under the transformation
$g_{\mu\nu} \to g_{\mu\nu}e^{2\la}$, with $\la=const$. The power
counting (\ref{omG-4der}) can be perfectly well applied in this
case, yielding $\om(G)+d(G) = 4$. This means that the theory is
also multiplicatively renormalizable. The disadvantage of this
model of gravity is that there is no automatic Einstein limit in
the low-energy domain. Let us remember that such a limit is
one of the main conditions of consistency of any model which
generalizes or modifies GR, so this situation should be seen as a
problem of the model.

We note, by passing, that one can start from such a theory
of globally conformal theory, coupled to a massless scalar field.
At the quantum level, the loop corrections to the scalar
potential may produce such effective potential that the global
conformal symmetry is dynamically broken, producing GR in
the low-energy limit. This idea resurrected several times in the
literature (see, e.g., \cite{BuchWeyl,induce,Agravity}) in
different QG frameworks and, in general, looks attractive and
promising.
Unfortunately, the real deal is that, in the IR one has to break the
global conformal symmetry. And then, in the broken phase, we
come back to the massive ghost problems and to the related
issue of instabilities, as it will be discussed in the subsequent
section~\ref{IIQG-chap2}.

Another particular case, which is  instructive to consider, is
the $R+R^2$-gravity, which is model  (\ref{action4der}) without the
$C^2$-term. As we saw in the previous sections, in this model the
traceless component of the metric ${\bar h}_{\mu\nu}$ has $1/k^2$
propagator, while the scalar mode has a propagator behaving as
$1/k^4$ in the UV.  Furthermore, there are vertices $K_{4d}$
connecting all these modes. It is easy to check that the power
counting in this model is dramatically different from that in the
general fourth-derivative model. The theory is non-renormalizable
and the power counting is even much worse than in quantum GR.

The next example is the model (\ref{action4der}) without the $R^2$
term. This particular model may be interesting since the
fourth-derivative part of the action possesses local conformal
symmetry. This symmetry is softly broken by the Einstein-Hilbert
and cosmological terms. The expression ``soft breaking'' means that
the symmetry does not hold in the terms with dimensional parameters.
Can it be that the softly broken
conformal symmetry ``saves'' the power counting in this model? The
answer is certainly negative.  The propagator of the traceless mode
of the metric, ${\bar h}_{\mu\nu}$, in this case, has the UV behavior
$1/k^4$, and that of the scalar mode the different UV behavior,
$\propto 1/k^2$, due to the presence of the $R$-term. At the
same time, there are $K_{4d}$ vertices that link all of the modes,
and hence the power counting is qualitatively the same as in the
previous $R+R^2$ case. The theory is not renormalizable.

The
situation is much more complicated if the original theory has the
pure Weyl-squared term in the Lagrangian, i.e., possesses local
conformal symmetry. The theoretical proof of the renormalizability
in conformal theories exists  only for the quantum theory in
curved space (semiclassical gravity) \cite{tmf84} and, in the
literature, there is no proof for conformal QG. On the other hand,
at the one-loop order this theory is renormalizable, as was shown
by direct calculations \cite{frts82,antmot} and confirmed in
\cite{Weyl}.
However, it is expected that this model is not renormalizable at
higher loops because of the conformal anomaly. But, in the
situation when this expectation is not supported by direct
higher-loop calculations or the analysis similar to \cite{tmf84},
the question should be regarded as open.

\subsection{Power counting in the polynomial theory}

\index{Power counting} 
Consider power counting in the polynomial model (\ref{gaction}).
As before, we assume that the Faddeev-Popov quantization is done
with the weight operator (\ref{weight-N}) and the correspondingly
modified ghost term (\ref{Mghost-mod}).
In the general case, both coefficients of the
highest derivative terms $\om_{N,R}$ and $\om_{N,C}$ are non-zero.
Then the propagators of both metric perturbations $h_{\mu\nu}$
and ghosts have the UV behavior $\propto k^{-4-2N}$, i.e., we have
$r_l \equiv 4+2N$. For the vertices,  the generalization of
Eq.~(\ref{nK-4der}) is
\beq
&&
\sum_VK_V\,=\,
n_{4+2N}K_{4+2N} +
n_{2+2N}K_{2+2N} +
n_{2N}K_{2N} + \,.\,.\,.\,
+ n_{EH} K_{EH} +
n_\La  K_\La\,,
\nonumber
\\
&&
V \,=\, n_{4+2N} + n_{2+2N} + n_{2N} + \,.\,.\,.\,
+ n_{EH}+n_\La,
\label{nK-Nder}
\eeq
where $K_{4+2N} = 4+2N$,  \ $K_{2+2N} = 2+2N$, \,.\,.\,.\,
$K_{EH} = 2$, \ $K_\La = 0$, \ and \ $n_{4+2N}$, $n_{2+2N}$,
\ $n_{2N}$, \,.\,.\,.\,, $n_{EH}$, and $n_\La$ are the numbers of
the respective vertices.

Consider the diagrams with the strongest divergences. This
means $V= n_{4+2N}$, such that the other types of vertices are
absent. Then the power counting becomes quite simple, because of
$\,\sum_VK_V= V(4+2N)$. The expression (\ref{omG-qg})
becomes
\beq
\om(G)+d(G)
&=&
(4 - 4 - 2N)I
\,-\,4V \,+\,4\, + \,V(4+2N)
\nonumber
\\
&&
\,\,=\,\,
4\, + \,2N(V-I)
\,=\,
4 \, + \,2N(1-L),
\label{omG-N}
\eeq
where we used the topological relation
 (\ref{topa}) in the form $V-I=1-L$. It is easy to see that the
 power counting in the four-derivative model (\ref{omG-4der})
 is a particular case, corresponding to $N=0$. Thus, we assume
 $N \geq 1$.

According to (\ref{omG-N}), the sum $\om(G)+d(G)$
decreases with a growing number of loops $L$.
The strongest divergences occur for $L=1$, when the
aforementioned sum equals 4 and the logarithmic divergences
correspond to the one-loop counterterms with, at most, four
derivatives. Taking the covariance and locality, this means that
the one-loop counterterms are of the  $C^2$, $R^2$, $E_4$ and
$\square R$ types. In other words, all the terms with six and more
derivatives do not need to be renormalized at the one-loop order.
But, if there just one vertex with two less derivatives,
i.e., $n_{2+2N}=1$ and $n_{4+2N}=V-1$, then we meet only
the Einstein-Hilbert type divergence.  And finally, in the case
$n_{2+2N}=2$ or $n_{2N}=1$, the unique divergence is that of
the cosmological constant.

The features of the $L=1$ approximation that we listed above, do not
depend on $N \geq 1$. Starting from $L \geq 2$, the structure of divergences
starts to depend on the value of $N$. In particular, for $N \geq 3$,
according to  (\ref{omG-N}), the second- and higher-loop diagrams
are all finite. This creates a situation when the one-loop
beta functions are the exact ones.  In the case of $N =2$, there
are two-loop divergences, but only of the cosmological constant
type. Finally, for $N =1$, there are two-loop divergences of the
Einstein-Hilbert and the cosmological constant type and also
three-loop divergences, but only of the cosmological constant type.

All in all, a theory with both $\om_{N,R}\neq 0$ and
$\om_{N,C}\neq 0$, is superrenormalizable. In contrast, in the
degenerate case, when only one of these coefficients is zero, the
theory is non-renormalizable. The situation is essentially similar to
what we discussed above for two similar four-derivative models
of QG.

Finally, the power counting in the QG models (\ref{act Phi}) with
non-polynomial (typically exponential) functions $\Phi$ and  $\Psi$
cannot be performed on the basis of the formula (\ref{omG-qg}),
because both the number of derivatives in the vertices and the
parameter $r_l$ are infinite. However, if the condition
(\ref{cond-nonloc}) is satisfied,  the evaluation can be done using
only the topological relation \cite{CountGhosts}. Let us assume, for
simplicity, that the functions are those from (\ref{PhiPsi}). Then,
in Euclidean signature, each propagator brings the exponential of
$-\al k^2$ and each vertex contributes with the exponential of
$+\al k^2$. This means, if the number of vertices is different from
the number of propagators, the last integral in the given diagram
will be either strongly divergent, or completely convergent. Taking
this into account, from the relation (\ref{topa}) we learn that the
divergences are present only in the one-loop diagrams and that these
divergences have fourth powers of momenta. Thus, the power
counting in the theories of this class is the same as in the
polynomial models with $N \geq 3$.

\section{Massive ghosts in higher-derivative models}
\label{IIQG-chap2}

Previously we saw that quantum GR, based on the Einstein-Hilbert
action, is a non-renormalizable theory. On the other hand, by adding
the general fourth-derivative terms, we arrive at the model providing
multiplicative renormalizability. Another strong argument for
including fourth-derivative terms is that they are required for the
renormalizability of the semiclassical theory, when gravity is an
external field \cite{birdav,book,ParToms,OUP}. And this is something
one cannot disregard. The point is that the concepts of QFT from one
side and of the curved space from another, are well-established and,
to a great extent, verified by experiments and observations. Thus, if
QFT in curved space produces the four-derivative terms in the action,
we have to admit that these terms are there. The problem concerns
not exactly UV divergences and formal renormalizability. Together
with the logarithmic divergences there are always logarithmic
nonlocal form factors. In the IR such a form factor behaves,
effectively, as a constant \cite{GWprT,OUP}. This means, if we do
not include fourth derivatives into the action, they will come there
anyway as legitimate corrections coming from quantum matter fields.
Therefore, independent on which approach to QG we choose, it makes
sense to include fourth derivatives terms into the gravitational
action and check out the physical consequences of this inclusion.

At the classical level and in the low-energy domain, the fourth
derivative terms look irrelevant because the coefficients of these
terms in the action (\ref{action4der}) are just a numbers while the
coefficient of the $R$-term is $1/G$, that is
$M_P^2 \approx 10^{38} GeV$. As a consequence, until the metric
derivatives, in the momentum representation, do not have Planck-order
frequencies, the fourth-derivative terms cannot compete with the
Einstein-Hilbert term. This situation is usually called a Planck
suppression. So, from the first sight the deal is perfect: including
the fourth derivatives terms we get a renormalizable QG (and
semiclassical too!), while all classical 
solutions remain the same as in GR and one can enjoy the
well-verified gravitational theory.

Unfortunately, even if theory (\ref{action4der}) is, formally,
multiplicatively renormalizable, it does not make it consistent at
either the quantum or even classical levels. As we shall see in brief,
the spectrum of this model includes states that have negative
kinetic energy. These states, or particles, are called massive
nonphysical ghosts. The presence of ghosts violates the unitarity of
the theory at the quantum level. Worse than that, in the presence
of massive ghosts or, more generally, ghost-like states, classical
solutions of the theory can be unstable with respect to the metric
perturbations. Qualitatively, the same situation takes place not
only in the fourth-derivative theory but in all polynomial models
of quantum gravity.

The problem of ghosts is certainly the main obstacle for building
a consistent QG theory. For this reason, we consider
this problem here. We shall closely follow \cite{OUP}. The interested
reader can address this book for more details, or go directly to the
original works, such as the reviews \cite{Woodard-review} on the
Ostrogradsky instabilities, or the papers \cite{HD-Stab} and
\cite{PP,Salvio-19,LinNonlin} for the approach we follow to
explore and interpret the instabilities caused by massive ghosts.

\subsection{What means a massive ghost}
\label{IIQG-sec2}

Consider the propagator of the transverse and traceless part of the
metric (\ref{Spin2-QG}) or the one of the scalar, gauge-invariant
mode (\ref{B2C-QG}). For simplicity, we can set
the cosmological constant to be zero, then the fourth-derivative
action (\ref{action4der}) becomes
\beq
S\,=\,-\,\int d^4x\sqrt{-g}\,
\Big\{\,\,\frac{1}{\la}\,R_{\mu\nu}^2
\,+\,\frac{\om - 1}{3\la}\, R^2  + \frac{1}{\kappa^2}\, R\,\Big\}.
\label{action-reduced}
\eeq
Formulas (\ref{Spin2-QG}) and (\ref{B2C-QG})
change accordingly, with $\Phi$ and $\Psi$ becoming constants.

We can consider at once both scalar and tensor modes,
because the formulas are similar. For the definiteness sake, consider
the tensor mode $\,h={\bar h}^{\bot\bot}_{\mu\nu}$, which is not
affected by the $R^2$ term. Using the expansions (\ref{R2-expand})
and (\ref{EH-expand}), the action of this mode becomes
\beq
S^{(2)}_{tensor}
= \int d^4x\Big\{
- \frac{1}{4\la}\big( \cx h \big)^2
-\frac{1}{4\ka^2} h \cx h \Big\}
= - \frac{1}{4\la}\,\int d^4x\,
h \big( \cx + m_2^2 \big) \cx h ,
\mbox{\quad}\mbox{\quad}
\label{action-tens}
\eeq
where $m_2^2 = \la/\ka^2$ is the mass of the mode that is
called a tensor ghost, a massive tensor ghost or higher-derivative
ghost. The reason for this exotic name is that the Euclidean
propagator of the spin-2 mode in this theory can be cast in the form
\beq
G_2 (k)
&\,\,\propto\,\,&
\frac{1}{m_2^2}\Big( \frac{1}{k^2}
\,-\, \frac{1}{k^2 + m_2^2} \Big)\,\,\hat{P}^{(2)}.
\label{propaHD}
\eeq
The negative sign of the second term indicates that the corresponding
mode is not a usual particle. In fact, we have not one but two degrees
of freedom of the tensor field. One of these degrees of freedom has
positive kinetic energy and zero mass, and it corresponds to the
first term in Eq.~(\ref{propaHD}). The second degree of freedom has
the mass $m_2$ and corresponds to the second term in
(\ref{propaHD}). As we shall see in what follows, its
kinetic energy is negative, and, for this reason, it is called a ghost.

The separation of the two degrees of freedom can be most
simply explored by using an auxiliary field $\Phi$ (see, e.g.,
\cite{Cremin}, a more detailed discussion in \cite{E.Alvarez-QG}
and more general formulations in \cite{fREddi}).
Consider the  Lagrangian density
\beq
{\mathcal L}'
&=&
-\,\frac{m_2^2}{4\la}\,h\cx h
\,+\, \la \phi^2 - \phi \cx h.
\label{actaux}
\eeq
The Lagrange equation for $\phi$ can be solved as
$\,\phi = \frac{1}{2\la} \cx h$. Substituting this expression back
into (\ref{actaux}), we arrive at Eq.~(\ref{action-tens}),
which shows the dynamical equivalence of the models
(\ref{action-tens}) and (\ref{actaux}).

The two fields $\Psi$ and $h$ in (\ref{actaux}) are not factorized.
To improve on this issue, we change to the new variables $\th$ and
$\psi$,
\beq
&&
h \,=\,\frac{ \sqrt{2\la}}{\,m_2}(a_1 \th + a_2 \psi),
\qquad
\phi \,=\, \frac{\sqrt{2\la}}{\,m_2}a_3 \psi ,
\label{aux1}
\eeq
where the unknown coefficients $a_{1,2,3}$ should provide the
separation of the modes and also the standard coefficients $\frac12$
or $-\frac12$ in the kinetic terms. A small algebra shows that the
condition $a_2+a_3=0$ is necessary to separate the variables and,
also, the condition $a_1=a_2=1$ is required to provide standard
normalization of the kinetic terms. Then, in the new variables, the
Lagrangian (\ref{actaux}) becomes
\beq
{\mathcal L}'
&=&
\frac{1}{2}\,\eta^{\mu\nu} \pa_\mu \th \pa_\nu \th
\,-\, \frac{1}{2}\,\Big(
\eta^{\mu\nu} \pa_\mu \psi \pa_\nu \psi
- m_2^2\psi^2\Big).
\label{aux3}
\eeq
We can
conclude that the theory (\ref{action4der}) has healthy tensor
massless degrees of freedom $\th$ and, on top of this, tensor
massive degrees of freedom $\psi$ with negative kinetic energy,
called nonphysical massive ghost.

\subsection{Classification of ghosts and tachyons}
\label{IIQG-sec2.1}

Consider a basic classification of ghosts and tachyons
following \cite{GWprT,OUP}.
The general action of a free second-order field
$\,h(x)=h(t,{\vec r})\,$ can be written as
\beq
S(h) &=&
\frac{s_1}{2} \int d^4x \,\big\{ \eta^{\mu\nu}\pa_\mu h \pa_\nu h
- s_2 m^2 h^2 \big\}
\nn
\\
&&
\quad  = \,\,
\frac{s_1}{2} \int d^4x \,\big\{ {\dot h}^2 - (\na h)^2
- s_2 m^2 h^2 \big\}\,.
\label{act-II}
\eeq
Here $s_1$ and $s_2$ are sign factors $\pm 1$
for different types of fields. In what follows, we consider all
four combinations of these signs.

It is useful to perform the Fourier transform in the space variables,
\beq
h(t,{\vec r}) \,=\,
\frac{1}{(2\pi)^3} \int d^3k \,
e^{i{\vec k}\cdot{\vec r}}\,h(t,{\vec k}).
\label{Fur}
\eeq
In a free theory, one can consider the dynamics of each component
$\,h\equiv h(t,{\vec k})\,$ separately. Such a dynamics is defined by
the action
\beq
S_{\vec k}(h) &\,=\,&
\frac{s_1}{2} \int dt \,\big\{ {\dot h}^2 - {\vec k}^2 h^2
- s_2 m^2 h^2 \big\}
\,=\,
\frac{s_1}{2} \int dt \,\big\{ {\dot h}^2
- m_k^2 h^2 \big\},
\mbox{\quad}
\label{act*}
\\
&&
\mbox{where}
\qquad
{\vec k}^2 = {\vec k}\cdot{\vec k}
\qquad
\mbox{and}
\qquad
m_k^2 = s_2 m^2 + {\vec k}^2 \,.
\label{kmass}
\eeq
The properties of the field are defined by the signs of $s_1$ and
$s_2$. The possible options can be classified as follows:

i) \ {\it Normal healthy field} \ corresponds to $s_1=s_2=1$. The
kinetic energy of the field is positive and the equation of motion
has the oscillatory form,
\beq
{\ddot h} + m_k^2 h \,=\,0\,,
\label{pp}
\eeq
with the usual periodic solutions.
\vskip 2mm

ii) \ {\it A tachyon} \ has $s_1=1$ and $s_2=-1$. The classical
dynamics of tachyons is described in the literature, e.g., in
\cite{Sudarshan,Terletsky}, but, for our purposes, it is sufficient
to give only a basic survey.
For a relatively small momentum $k=\vert {\vec k}\vert $, there is
$m_k^2 < 0$ in Eq.~(\ref{kmass}), and the equation of motion is
\beq
{\ddot h} - \om^2 h \,=\,0\,,
\qquad
\om^2 = \left|m_k^2\right|\,,
\label{pm}
\eeq
with exponential solutions
\beq
h\,=\,h_1 e^{\om t}\,+\,h_2 e^{-\om t}\,.
\label{om}
\eeq
However, if such a particle moves faster than light, the solution
is of the normal oscillatory kind, indicating
that such a motion is ``natural'' for this kind of particle.

iii)
\ {\it A massive ghost} \
has $s_1=-1$ and $s_2=1$. It is not a tachyon, because $m_k^2\geq 0$.
In this case, the kinetic energy of the field is negative, but the
Lagrange equation (leaving aside its derivation from the least action
principle) has a normal oscillatory equation (\ref{pp}).

iv)
\ {\it A tachyonic ghost} \
has $\,s_1=s_2=-1$. For relatively small ${\vec k}^2$, one meets
\ $m_k^2 < 0$. The kinetic energy is negative and the mass is
imaginary. So, along with the problems typical for the ghosts,
the free wave solutions are exponential, as in (\ref{om}). One
can find a discussion of the implication of tachyonic ghosts
in \cite{GWprT}.

\subsection{Massive ghosts in the fourth-order model}
\label{IIQG-sec2.1.2}

Let us come back to the fourth-order gravity model
(\ref{action-reduced}). Consider the free tensor modes
in the theory (\ref{action-tens}) and change to the
momentum representation,
\beq
\cx h \, \, \,=\, \, \, {\ddot h} - \De h
\, \, \,\longrightarrow\, \, \,
{\ddot h} + {\vec k}^2 h .
\eeq
Here ${\vec k}$ is the wave vector of an individual mode.
It is important that, owing to the presence of both massless and
massive modes, the standard massless dispersion relation between
the frequency and the wave vector does not hold in this case. The
Lagrange function of the wave with fixed ${\vec k}$ can be obtained
from (\ref{action-tens}):
\beq
L
&=&
-\,\frac{1}{4\la}\,\big( {\ddot h} + {\vec k}^2 h \big)^2
\,-\,
\frac{1}{4\ka^2}\,h \big( {\ddot h} + {\vec k}^2 h \big).
\label{Lk}
\eeq
The Lagrange equation for $L=L(q,\,{\dot q},\,{\ddot q})$ has
the form
\beq
\frac{\pa L}{\pa q}
\,-\,
\frac{d}{d t} \frac{\pa L}{\pa {\dot q}}
\,+\,
\frac{d^2}{d t^2} \frac{\pa L}{\pa {\ddot q}}
\,=\,0
\label{Leqs-4d}
\eeq
and the energy can be easily obtained in the form
\beq
E
\,=\,
{\dot q}\, \Big(
\frac{\pa L}{\pa {\dot q}}
\,-\, \frac{d}{d t}\frac{\pa L}{\pa {\ddot q}}
\Big)
\,+\,
{\ddot q}\, \frac{\pa L}{\pa {\ddot q}}
\,-\,L\,.
\label{E-4d}
\eeq
In our case (\ref{Lk}), this formula gives
the energy of the individual wave with momentum ${\vec k}$,
\beq
E
\,=\,
\frac{1}{4\la}\,\Big( 2 h^{\lp\textsc{\tiny III}\rp} {\dot h}
- {\ddot h}^2 \Big)
\,+\, \Big(
\frac{1}{4\ka^2} + \frac{{\vec k}^2}{2\la} \Big)\,{\dot h}^2
\,+\,
\Big(\frac{{\vec k}^2}{4\ka^2} + \frac{{\vec k}^4}{4\la} \Big)
\,h^2\,.
\label{Ek}
\eeq
This formula provides some information about the
fourth-derivative theory. We can separate it into the following
points:

i) \ In the limit $\la \to \infty$, the remaining expression for the
energy is positively defined, as it should be for Einstein's gravity.

ii) \ The fourth time derivative terms are given by the first
summand in (\ref{Ek}). It is easy to see that this term is not
positively defined. This sign indefiniteness should be expected,
as a direct consequence of the presence of the massive nonphysical
ghost.

iii) \   In the model under discussion, the low-energy limit
(IR) means
\beq
{\ddot h}^2 \ll {\vec k}^2 {\dot h}^2
\qquad
\mbox{and}
\qquad
\vert h h^{\lp\textsc{\tiny III}\rp}
\vert \ll {\vec k}^2 {\dot h}^2.
\label{IR-QG}
\eeq
In this case, the first indefinite term (with fourth derivatives) in
(\ref{Ek}) is small, and the sign of the energy is defined by the
second term, providing a relevant constraint on the action
(\ref{action4der}). The positivity of the theory in this limit does
not depend on fourth time derivatives. However, the kinetic energy
can be still unbounded from below for the negative coupling
$\la < 0$. Owing to
the violated dispersion relation between the wave vector ${\vec k}$
and the time derivatives, it is possible to have a large ${\vec k}^2$
with the conditions (\ref{IR-QG}) satisfied. Thus, the sign of the
coupling $\la$ in the action (\ref{action4der}) should be positive,
as it was always assumed in the literature, e.g., in the classical
works \cite{Stelle77,stelle78} and \cite{frts82}.

The equation for tensor perturbations can be derived from
(\ref{Leqs-4d}),\footnote{The more general
equation
describing the dynamics of tensor perturbations on the cosmological
background, will be discussed below; see Eq.~(\ref{pertu}).}
\beq
h^{\lp\textsc{\tiny IV}\rp}
+ 2{\vec k}^2{\ddot h}  + {\vec k}^4 h
\,+\,
\frac{\la}{16 \pi \ka^2}\, \big({\ddot h} + {\vec k}^2 h \big)\,=\,0.
\label{hhhh}
\eeq
One can introduce the new notation,
\beq
\frac{\la}{16\pi \ka^2} = s_2 m^2,
\label{m2}
\eeq
where $\,s_2 = {\rm sign}\,\la\,$ and $\,m^2>0$.
Then Eq. (\ref{hhhh}) becomes
\beq
\Big( \frac{\pa^2}{\pa t^2} + {\vec k}^2 \Big)
\,\Big( \frac{\pa^2}{\pa t^2} + m_{k}^2 \Big)h\,=\,0\,,
\label{factor}
\eeq
where $m_k^2$ was defined in (\ref{kmass}).
The solutions of the last equation can be different, depending
on the sign of $\la$ and, hence, that of $s_2$.  The general
formula for the frequencies in $h \sim \exp{\{\pm \om t\}}$ has the form
\beq
\om_{1,2} &\approx&
\pm\,i\big({\vec k}^2\big)^{1/2}
\qquad
\mbox{and}
\qquad
\om_{3,4} \approx
\pm \big(-\,m_k^2\big)^{-1/2}\,.
\label{osc}
\eeq
For a positive $\la$, there are only imaginary $\om's$ and, hence,
oscillator-type solutions. In contrast, for $\la < 0$, we have
$s_2=-1$ and the roots $\om_{3,4}$ are real, since, in this
case, $-m_{k}^2>0$ for sufficiently small ${\vec k}^2$.
Indeed, the first couple of roots corresponds to the massless
graviton, and the second couple to the massive particle. According
to the classification presented above, this particle is a ghost for
$\,\la > 0$ and it is a tachyonic ghost for $\la < 0$.

Finally, we conclude that the model (\ref{action4der}) has ghosts
(and maybe tachyonic ghosts) owing to the presence of fourth
derivatives.

\subsection{Massive ghosts in the six and higher-order models}
\label{IIQG-sec2.1.6}

If the number of derivatives in the polynomial model of six or
greater, the structure of the propagator is more complicated than
in the fourth-order theory. First of all, the massive poles can be
either real or complex. In the last case, they emerge in the complex
conjugate pairs. There is an important theorem about the structure
of the propagator in the case of real poles \cite{highderi}. In this
case, instead of (\ref{propaHD}), we meet
\beq
G_2 (k)
\,=\,
\frac{A_{0}}{k^2} +  \frac{A_{1}}{k^2 + m_1^2}
+ \frac{A_{2}}{k^2 + m_2^2}
+  \,.\,.\,.\,  + \frac{A_{N+1}}{k^2 + m_{N+1}^2},
\label{28}%
\eeq
where  the squares of the masses $m_j^2$ are real. Assuming that
there is the following hierarchy of the masses:
\beq
0 < m_1^2 < m_2^2 < m_3^2 <  \,.\,.\,.\,  < m_{N+1}^2,
\label{orderma}
\eeq
one can prove that the signs of the coefficients $A_j$ alternate,
i.e., $\sign\ [A_j] = -\; \sign\ [A_{j+1}]$. This feature means one
cannot choose the theory in a such a way that all degrees of
freedom instead of the heaviest one are normal particles and
the mass of the ghost is infinitely large.

The complex poles are also possible, but their detailed discussion
lies beyond the scope of this review. Interesting hints about the
role of real and complex poles come from the analysis of the
Newtonian limit and the bending of light in the polynomial models
\cite{Modesto2014,Giacchini2016,ABSh2}.

\subsection{On the quantum consistency and stability of classical
solutions}
\label{IIQG-sec2.1.3}

We have seen that the theory with fourth derivatives, typically, has
a massive ghost or a tachyonic ghost, and this conclusion can be
extended to the polynomial models with more than four derivatives.
So, it is interesting to understand what the ghost means, from the
physical viewpoint. Let us give just a brief qualitative description
of the situation.

A particle with negative kinetic energy tends to the minimum of
the action and therefore tends to achieve a maximal speed. If such
a particle is free, it cannot accelerate, as this would violate
energy conservation. Hence, a free ghost does not produce any
harm to the environment, being isolated from it.
However, in the case when there is an interaction of a ghost with
healthy fields, the argument about energy conservation in a closed
system does not work.  Since any physical system tends to the
state with minimal action, a ghost tends to accelerate, transmitting
the extra positive energy to the healthy fields interacting with it,
in the form of the quantum or classical emission of the
corresponding particles. A systematic study of this situation at the
quantum level has been given by Veltman in \cite{Veltman-63}.

Since gravitons are massless and the metric-dependent gravitational
theories have non-polynomial interactions, a massive  ghost always
couples to an infinite amount of gravitons. This fact may lead to
dramatic consequences. The energy conservation does not forbid
a spontaneous creation of a massive ghost from the vacuum, even
in the flat Minkowski space. It is clear that such a spontaneous
creation of a ghost also implies that the corresponding amount of
positive energy should be released with the creation of massless
gravitons. As the mass of the ghost has the Planck magnitude,
these gravitons have to accumulate with the Planck energy density.

Assuming the existence of even a single real ghost, such a particle
should accelerate, emitting and scattering gravitons. The magnitude
of the energy of the ghost would increase, and hence the energy of
the created and scattered gravitons would increase too, without an
upper bound for the emitted gravitational energy. After a while, the
ghost would acquire an infinite amount of negative energy and start
to emit an infinite amount of positive energy. It is clear that if some
objects of this sort would be around, we would certainly know about
it or, rather, we would feel it.

Thus, the main theoretical problem is to explain why this dramatic
scenario does not work. We have to say that, at the moment, there
is no solution to this important problem. The solution could be,
e.g., an explanation of why gravitons cannot agglomerate with the
Planck density, but no mechanism for this has been formulated.
Obviously, we have to assume that some kind of solution exists.
Probably, it is related to the large mass of the massive ghost in
higher-derivative gravity.

The simplest way to preserve the
unitarity is to admit the existence of a ghost. But, as we described
above, this leads to the physically inconsistent output.  Thus, one
avoids ghosts by forming the \textit{in} states only with gravitons.
Owing to the interactions, ghost wakes up from the vacuum and
emerge in the \textit{out} states -- that means the scattering matrix
is not unitary \cite{Stelle77} and we arrive at the contradiction.

Historically, the main efforts in solving the problem of ghosts was
related to the quantum aspects. In this respect we can start the
list from the mentioned paper by Veltman \cite{Veltman-63},
regardless it has nothing specific about quantum gravity. Soon
after the seminal work of Stelle \cite{Stelle77} with the proof of
renormalizability of the fourth-derivative QG, there were first
works about solving the problem of ghosts \cite{Tomboulis-77} and
\cite{salstr}. The main common idea of these works was that the loop
corrections to the propagator (\ref{propaHD}) transform the unique
massive pole into a pair of the complex conjugate poles, with the
positions of these poles being gauge fixing dependent \cite{antomb}.
In this case, one can
prove the unitarity of the $S$-matrix, violated by the presence of
massive ghosts \cite{Stelle77}. Unfortunately, it was shown
\cite{Johnston}, that definite conclusion on this issue can be taken
only on the basis of \text{exact} knowledge of the dressed
propagator of $h_{\mu\nu}$. E.g., the one-loop corrections or
the $1/N$ approximation are insufficient for solving the problem.
It is interesting that starting from the six-derivative theory, one
can provide the desirable features (pair of complex conjugate
poles, for instance) already at the tree-level and, also, prove that
the loop corrections do not change this structure. In this
situation, one can use the optical theorem and prove the unitarity
of the $S$-matrix in the Lee-Wick approach \cite{Modesto-complex}.

The problem with this solution of the ghost problem is that the
Lee-Wick approach assumes that the scattering occurs between
the asymptotical states, where \textit{in} and \textit{out} states
describe the free particles. However, in gravity (and especially in
its quantum version) the notion of free massive particle is not
perfectly well defined, because any such particle produces a
gravitational field, starts to interact with it and, therefore, is not
completely free. Therefore, the definite resolution of the ghost
problem by means of the $S$-matrix does not look really
promising, at least at the fundamental level.

What we can learn from all the quantum considerations is that the
issues with ghosts described above would become impossible in the
presence of a natural cut-off on the energy density of gravitons,
such that this density never achieves the Planck order of magnitude.
Then the gravitons cannot agglomerate to create a ghost and the
$S$-matrix remains unitary. The problem is that there is no
theoretical mechanism for such a cut-off. One can say that this is
the main reason of why the problem of ghosts does not have a
solution.

\subsection{Stable solutions in the presence of massive ghosts}
\label{IIQG-sec2.5}

Another aspect of the problem of ghosts is related to the stability
of classical solutions.
The most important issue related to massive ghosts is whether
their presence can be compatible with the stability of the classical
solutions of GR. As we mentioned above, in higher derivative models
of gravity one typically meets the Planck suppression. As a result,
classical solutions of GR represent high-quality approximations to
the corresponding solutions in the presence of the higher-derivative
terms. This logic can be successfully applied to the fourth-derivative
theory and extended to the polynomial theories (\ref{gaction}), if
we assume that all massive parameters are of the Planck order
of magnitude \cite{ABSh2}.

The excellence in the approximation of the solution
of GR does not necessarily mean the stability of this solution. In
general, providing the stability of a gravitational solution under
arbitrary small perturbations (that do not have the symmetry
of solutions themselves) may be not simple even in GR, for some
gravitational backgrounds. Let us mention, for instance, the study
of the stability of the Schwarzschild solution in GR \cite{RW,Zer}
(see also \cite{FrolovNovikov}).
In the presence of $C^2$ term, one can expect that the same solution
will not be stable. Regardless of existing contradictions in the
literature, in general, this expectation is confirmed \cite{whitt,myung}.
Owing to the high level of technical difficulty, this case will not be
discussed here.

From the technical side, it is much simpler to consider the stability
of classical cosmological solutions in the presence of fourth-order
terms. The advantage of this simpler case is that the physical
interpretation of the results is relatively explicit. The analysis of
\cite{HD-Stab} and after that in \cite{PP,Salvio-19,LinNonlin},
was done for the fourth-derivative  action (\ref{action4der}) with
anomaly-induced semiclassical corrections (see, e.g., \cite{OUP}
for the review). It turns out that these corrections do not change
the main result. We shall explain this result omitting all technical
details except the basic formulas.

The stability we need to explore is related to the presence of
massive spin-2 ghost degrees of freedom, which means the transverse
and traceless modes of the metric perturbation on the homogeneous
and isotropic, cosmological background.
According to the theory of cosmological perturbations
(see, e.g., \cite{Dodelson,Mukhanov,RubakovGorbunov}), the
background cosmological metric with tensor perturbations is
\beq
\dd s^2\,=\,
a^2(\eta) \big[ \dd\eta^2 -  \lp\ga_{ij} + h_{ij}\rp \dd x^i
\dd x^j \big],
\label{expans}
\eeq
where $\eta$ is the conformal time, $a(\eta)$ corresponds to a
background cosmological solution, and we imposed the synchronous
coordinate condition $h_{\mu0} = 0$. Furthermore, for the sake of
simplicity we consider the flat space geometry $k=0$ (hence,
$\ga_{ij}=\eta_{ij}$) and set the cosmological constant to vanish,
$\La=0$.

Since we are interested in the gravitational wave
dynamics, it is sufficient to retain only the traceless and transverse
parts of $h_{ij}$, which are the purely tensor modes, by imposing
\beq
\pa_i\,h^{ij}=0
\,,\qquad
h_{kk}=0 \,.
\label{spin2}
\eeq
As before, we do not need to write indices and set $h=h^{ij}$.

The Lagrange equation for $h(t)$, in terms of physical time,
has the form  \cite{GW-Stab}
\beq
&&
\frac{1}{3}  h^{\lp\textsc{\tiny IV}\rp}
+ 2H h^{\lp\textsc{\tiny III}\rp}
+ \lp H^2 - \frac{\la\MP^2}{16\pi}\rp \ddot{h}
+ \frac{2}{3} \lp \frac14 \frac{\nabla^4 h}{a^{4}}
- \frac{\nabla^2 \ddot{h}}{a^2}
-  H \frac{\nabla^{2} \dot{h}}{a^{2}}\rp
\nonumber
\\
&&
\quad
- \lp H \dot{H} + \ddot{H} + 6 H^3
 + \frac{3\la\MP^2 H}{16\pi}\rp \stackrel{.}{h}
+ \lc \frac{\la\MP^2}{16\pi }
+ \frac43  \lp \dot{H} + 2H^2\rp
\rc \frac{\nabla^2 h}{a^{2}}
\nonumber
\\
&&
\quad
- \lc
24 \dot{H} H^{2} + 12 \dot{H}^{2} + 16 H \ddot{H}
+ \frac83 H^{\lp\textsc{\tiny III}\rp}
+ \frac{\la\MP^{2}}{8\pi}\lp 2 \dot{H} + 3 H^{2}\rp\rc h = 0\,.
\mbox{\qquad}
\mbox{\quad}
\label{pertu}
\eeq
The contribution of the fourth-derivative terms depends on the unique
parameter $\la$ from the action (\ref{action4der}). The reason is
that the Gauss-Bonnet combinations do not affect the equations
of motion and another invariant is $R^2$, which contributes to the
equation for the conformal factor $a(t)$, but does not affect the
propagation of the tensor mode.

The next step is to make the Fourier transformation for the
space coordinates
\beq
h_{\mu\nu}({\vec r},t)
&=&
\int \,\frac{d^3k}{(2\pi)^3}\,\,
h_{\mu\nu}({\vec k},t) \,\,e^{i{\vec k}\cdot{\vec r}} .
\label{Fu}
\eeq
One can treat the wave vector ${\vec k}$
as a constant and hence will be interested only in the time evolution
of the perturbation $h_{\mu\nu}({\vec k},t)$. The validity of such a
treatment is restricted to the linear perturbations, but this is
what we need now. In this way, the complicated partial differential
equation (\ref{pertu}) are reduced  to the much simpler ordinary
differential equation for each of the individual modes.

Using the notation $h=h(t,{\vec k})=h(t,k)$, the equation has the form
\beq
&&
h^{\lp\textsc{\tiny IV}\rp}
+ 6H h^{\lp\textsc{\tiny III}\rp}
+ \lp 3H^2 - \frac{3\la\MP^2}{16\pi}\rp \ddot{h}
+ \lp \frac12 \frac{k^4}{a^{4}}h
+ \frac{2k^2 }{a^2}\ddot{h}
+  \frac{2k^2}{a^2}\,H\dot{h}\rp
\nonumber
\\
&&
\quad
-\, 3 \lp H \dot{H} + \ddot{H} + 6 H^3
 + \frac{3\la\MP^2 H}{16\pi}\rp \stackrel{.}{h}
- \lc \frac{3\la\MP^2}{16\pi }
+ 4 \lp \dot{H} + 2H^2\rp
\rc \frac{k^2 }{a^{2}}h
\nonumber
\\
&&
\quad
-\, \lc
72 \dot{H} H^{2} + 36 \dot{H}^{2} + 48 H \ddot{H}
+ 8 H^{\lp\textsc{\tiny III}\rp}
+ \frac{3\la\MP^{2}}{8\pi}\lp 2 \dot{H} + 3 H^{2}\rp\rc h = 0\,,
\qquad\qquad
\label{pertu-k}
\eeq
where $k=|\vec{k}|$ is the frequency of the massless field.
%
Finally, the initial conditions for the perturbations will be chosen
according to the quantum fluctuations of free fields. The spectrum
is identical to that of a scalar quantum field in Minkowski space
\cite{birdav},
\beq
h(x,\eta) = h(\eta)\,e^{\pm i\vec k.\vec r} \,\,,\,\,
\qquad
h(\eta) \propto \frac{e^{\pm ik\eta}}{\sqrt{2k}} \,.
\label{incon}
\eeq
As before, $\eta$ is conformal time. The transition to the physical
time is $a(\eta)d\eta = dt$. The normalization
constant in (\ref{incon}) is irrelevant for the linear perturbations.

The possible instabilities can
be explored using Eq.~(\ref{pertu-k}). According to the known
mathematical theorems about the stability of the fixed points of
differential equations, linear stability guarantees non-linear (at
least perturbative) stability for sufficiently small perturbations.
In the present case, this point was confirmed in
\cite{Salvio-19,LinNonlin} for the Bianchi-I metric, that is a
reduced form of the perturbation for ${\vec k}=0$.

The main qualitative result of the numerical analysis of
Eq.~(\ref{pertu-k}) performed in \cite{HD-Stab,PP} is that
the linear stability in the fourth-derivative model is possible,
but the two conditions should be fulfilled.
First of all, the frequency $k$ should be essentially smaller than the
Planck mass. The threshold value for $k$ slightly depends on the
type of the cosmological solution  (dominated by radiation, dust
or cosmological constant), but, in
all cases for  $k < 0.1\,M_P$, there is no growth of $h(t,k)$,
while such a growth is evident starting from $k \approx 0.6\,M_P$.
This requirement exactly corresponds to our expectation that one
needs a Planck density of gravitons to wake up the ghost from the
vacuum.  For a small frequency $k$, the ghost remains as a virtual
mode and cannot be created from vacuum to become a real particle.
Let us remember that the creation of a ghost from the vacuum
requires positive-energy gravitons with the Planck energy density
(Planck energy in the space volume of a cube of the Planck-scale
Compton wavelength). If the frequency of the gravitational wave is
insufficient, the ghost is not created, and there is no instability.

Second, the signs of both parameters $\la$ and  $\ka^2$ should
be positive. For a negative sign of the calibrated Newton constant
$\ka^2$, the ghost  becomes also a tachyon and there is no stability
for any frequency. The negative sign of $\la$ means the graviton
becomes a ghost and the massive particle is normal. Then there
is no threshold for creating a ghost and we observe instabilities
at all frequencies. And this is exactly what was observed in the
numerical analysis in \cite{HD-Stab}. It worth noting that, in the
previous subsection, we saw that $\la>0$ is a condition of stability
in the flat spacetime. Now we can see that this is confirmed by the
analysis of stability on the cosmological background. Thus, in
what follows we assume that $\la$ and $\ka^2$ are both positive.

As we mentioned already a few times, the solution of the ghost
problem (and, consequently, of the QG problem in general)
requires an explanation of why gravitons cannot accumulate
with the Planck energy density.\footnote{One can find the
discussions of other implications of this unknown physical
principle in \cite{Dvali}.} Intuitively, it is easier to
accept that such accumulation may occur only when the
background metric describes an intensive gravitational field,
as in the early Universe.

In this respect, an interesting thing happens when the frequency
$k$ is greater than the energy threshold but the external
cosmological background is described by a strong gravity. The
last means that the Hubble parameter has a large value and the
Universe is rapidly expanding. It turns out that there is a very
fast growth of $h(t,k)$ but such an explosion of the perturbations
does not last for a long period of time. To understand why this
happens, one can take a look at the main equation, (\ref{pertu-k}).
The frequency $k$ enters this equation in the combination $q=k/a(t)$.
For a sufficiently fast expansion of the Universe, the explosive
growth of the perturbation lasts only until the magnitude of $q$
becomes smaller than the energy threshold. After that, the
amplitude of the perturbations vanish exponentially. Thus,
the perturbations do not violate the cosmological principle, i.e.,
the Universe remains homogeneous and isotropic at the large
scale and the effect of ghosts does not contradict the observations.

To end this subsection, let us stress that the result of
\cite{HD-Stab,PP} and \cite{Salvio-19,LinNonlin} cannot be
interpreted as a solution of the problem of massive ghosts.
In our opinion, it  should be regarded as a hint to the direction
where such a solution can be found.

\subsection{Effective approach to the problem of ghosts}
\label{IIQG-sec2.4}

The effective approach to QG is a subject of a special Section of
 our Handbook, so it makes no sense to go into details of this
 approach here. Let us just briefly explain what is conventionally
 understood as an effective solution to the problem of massive
 ghosts, as introduced by Simon in \cite{Simon-90} and
 elaborated further in  \cite{ParSim}.

The proposal in  \cite{Simon-90} is to consider the Einstein
equations as the basic gravitational theory, regarding all
higher-derivative terms in the gravitational action and the
respective dynamical equations as a small perturbation.
According to this treatment, the gravitational theory should be
described by the two physical degrees of freedom of GR by
definition -- independent of what the action of the theory is
and the form of the quantum corrections to this action. The
propagator of the quantum metric $h_{\mu\nu}$ is derived from
Einstein's gravity, and no corrections can produce additional poles
in this propagator. By construction, there cannot be any kind
of massive ghosts, and hence there are no problems with unitarity
and instabilities at the classical or quantum levels.

This solution certainly looks mathematically correct and efficient.
On the other hand, there are serious problems with its consistency,
especially at the high energy scale, where the fourth-derivative
terms gain the magnitude comparable to the GR action. On top
of that, there is a problem with the uniqueness of the procedure.
E.g., one can modify this effective approach to include an $R^2$
term, or any $f(R)$ term, into the main part of the action,
because these terms do not produce a ghost. The same concerns
many other terms, e.g., all $\mathcal{O}(R^3_...)$'s. At the same
time, it should be strictly forbidden to do the same with the
$R\cx R$-term, which produces a scalar ghost. As it was discussed
in \cite{ABSh2} and also in \cite{OUP}, the analogy with QED and
the standard resolution of the problem of ``run-away solutions'' is
not convincing. So, all the scheme looks as an \textit{ad hoc}
procedure, without the physical background. It is as saying that we
do not like ghosts and will therefore forbid them. If we follow the
same approach in other branches of QFT, it is perfectly well
possible to modify any theory in a way we like and provide the
predictions we like. Despite this may look a universal solution of
all the problems, the theories created in this way would not be
reliable or, better say, would not provide reliable predictions.

However, if we restrict the area of application of QG to the
energies essentially lower that the Planck threshold, the described
effective approach becomes a normal feature of the theory that can
be fixed only by the observations and/or experiments. In this case,
the approach of \cite{Simon-90,ParSim} becomes equivalent to the
one described in the previous subsection. The results on the stability
of the cosmological background of \cite{HD-Stab} show that, in the
IR, one can trade using GR as a basic theory ``by definition'' to the
restrictions on the initial seeds of the tensor mode of cosmological
perturbations.

To conclude this section, we have to say that the problem of ghosts
is unsolved, at least if we do not restrict the applicability of QG
to the low-energy (IR) domain.  However, there are certain glues
about the directions in which the solution may be found. It seems
that we need a new physical principle forbidding the concentration
of gravitons with the Planck densities. This means, at the Planck
frequencies gravity should dramatically change. Such a change may
be because of the nonlocalities in the action, but the problem may
require a more complicated solution. At the UV, ghosts may be
generated from the vacuum. In the preprint \cite{Whois}, Hawking
made a hypothesis, that in this situation, the QFT approach should
be modified, taking into account that the ghost is not an independent
particle, but appears paired with the graviton. Thinking along this
line, one can expect to find a solution by working with bound states
or condensates, including ghosts and maybe some normal degrees
of freedom, e.g., in the framework of the superrenormalizable QG.

\section{Gauge-fixing dependence using general formalism}
\label{secQG1.5.2}

Starting from this point, in this and the next sections, we discuss
the loop corrections in the models of QG. This is an extensive
subject and it is traditionally one of the most worked out parts of
QG. Obviously, all of it cannot be settled into a short review, so
we shall discuss only two particular aspects. Namely, we perform
an important general analysis of the gauge- and parametrization
dependence of the loop corrections in QG, and show the derivation
of loop corrections in the simplest case of quantum GR in the
simplest gauge and parametrization of quantum metric field, i.e.,
repeat the main part of the paper by 't Hooft and Veltman \cite{hove}.
Another Chapter of this Section is devoted to the divergences in the
fourth-derivative QG model.

In this section, we show how the general statement about the on shell
gauge-fixing and parametrization independence of the effective action
can be used in different models of QG. The practical applications
described below, were introduced in \cite{frts82} and later on,
formulated in a more explicit form in \cite{avramidi-86} and \cite{a}.

\subsection{Gauge-fixing dependence in quantum GR}
\label{IIQG-sec3.1}

Consider the gauge-fixing and parametrization dependence in QG
based on GR. We consider the non-zero cosmological constant
for generality and also because an interesting application to the
on-shell renormalization group \cite{frts82}.

Let us start from some historic note and references. The subject
was pioneered in the paper \cite{KTT}, where the calculations
for the two-parameters gauge were performed using Feynman
diagrams. Even more general diagrams-based calculation in the
non-minimal gauge were done  in \cite{Kalm}. The use of
the heat-kernel methods required the generalized Schwinger-DeWitt
technique \cite{bavi85}, the results were applied to quantum GR
in \cite{DalvitMazz}.  On the other hand, in \cite{FirPei} it was
noted that one can simplify things by exploring the parametrization
ambiguity instead of the minimal gauge fixing.
The most general version of such a calculation \cite{JDG-QG},
used the background field method with the parametrization
\beq
\n{bgf}
g_{\al\be}
&
\,\,\longrightarrow \,\, & g'_{\al\be}
= e^{2 \ka r\si}
\Big[
g_{\al\be}
+ \ka \big(\ga_1\, \phi_{\al\be}
+ \ga_2\, \phi\, g_{\al\be} \big)
\nonumber
\\
&&
\,+\,\ka^2 \big(\ga_3\, \phi_{\al\rho}\phi^\rho_\be
+ \ga_4\, \phi_{\rho\om} \phi^{\rho\om} \,g_{\al\be}
+ \ga_5\,\phi \, \phi_{\al\be}
+ \ga_6\, \phi^2 \,g_{\al\be}
\big) \Big],
\eeq
where $g_{\al\be}$ is the background metric and $\phi_{\al\be}$ and
$\si$ are the quantum fields. Furthermore, the trace is defined as
$\phi = \phi^\mu_\mu$ and the indexes are lowered and raised with
the background metric $g_{\al\be}$ and with its inverse $g^{\al\be}$.
For the one loop effects, Eq.~(\ref{bgf}) is generalized
version of the simplest parametrization (\ref{bfm}), as it includes
arbitrary coefficients $\ga_{1,2,\dots,6}$ and $r$, which
parameterize the choice of the quantum variables. An important
detail is that, since the one-loop divergences are defined only
by the bilinear in quantum fields part of the action, (\ref{bgf}) is
the most general parametrization at the one-loop order. The next
point is that, for any choice of $\ga_{1,2,\dots,6}$ and $r$, one can
choose the gauge fixing parameters $\al$ and $\be$ such that
the bilinear form of the total action with the $S_{gf}$-term
(\ref{Sgf}), be a minimal operator. This approach enables one to
verify the general statements about gauge-fixing and parametrization
dependence in a relatively economic way, avoiding working with the
nonminimal operators.

In this review, we will not go into details of the practical
calculations, which can be seen in the mentioned original works,
but follow \cite{a,JDG-QG} and \cite{OUP} to explore all the
mentioned dependencies in the general framework.

The classical equations of motion corresponding to the
Einstein-Hilbert action with the cosmological constant term
(\ref{EH}), are (we ignore the irrelevant factor of $\ka^2$)
\beq
\vp^{\mu\nu}\,=\,\frac{1}{\sqrt{-g}}\,\frac{\de S}{\de g_{\mu\nu}}
\,=\,R^{\mu\nu}-\frac12\,\big(R+2\La\big)g^{\mu\nu}\,.
\label{vp}
\eeq
The general statement about gauge-fixing and parametrization
independence on-shell can be used together with the locality
of the divergent part of the effective action.  The power counting
tells us that this divergence has the form
\beq
\Ga^{(1)}_{div} &=&\frac{1}{\ep}\,
\int d^4x\sqrt{-g}\big\{
c_1\,R_{\mu\nu\al\be}^2 + c_2R_{\al\be}^2 + c_3R^2
+ c_4  {\Box}R + c_5R + c_6 \big \},
\qquad
\label{1-loop}
\eeq
where $\,\ep=(4\pi)^2(n-4)\,$ is the regularization parameter and
$c_{1,2,...,6}$ are some coefficients. Our purpose is to explore how
these coefficients depend on the parametrization and gauge fixing
choices.

We denote $\al_i$ the full set of arbitrary parameters
characterizing the gauge fixing and parametrization of the quantum
metric. The special values $\al^{0}_i$ of these parameters
correspond to  some fixed choice, e.g., to those in \cite{hove}.
The $\al_i$-related ambiguities in $\Ga^{(1)}_{div}$ do not
violate the locality of this expression. Taking this into account,
the on-shell universality tells us that
\beq
\de \Ga^{(1)}_{div}
&=&
\Ga^{(1)}_{div}(\al_i) \,-\, \Ga^{(1)}_{div}(\al^{0}_i)
\label{amb}
\\
&&
=\, \frac{1}{\ep}\,
\int d^4x\sqrt{-g}\,\big(
b_1 R_{\mu\nu} + b_2 Rg_{\mu\nu} + b_3  \La g_{\mu\nu}
+ b_4 g_{\mu\nu}\Box + b_5 \na_\mu \na_\nu\big) \,\vp^{\mu\nu},
\nn
\eeq
where the new parameters
$b_{1,2,..,5}$ in (\ref{amb}) depend on  $\al_i$ and
the explicit form of the dependence can be seen only from the
real calculations. However, one can draw relevant conclusions
directly from (\ref{amb}). In the simplest case of $\La=0$, this
formula tells us that only the Gauss-Bonnet counterterm $\int E_4$
cannot be set to zero by choosing $\al_i$. This is exactly the result
that was discovered by direct calculation in \cite{KTT}. The
$S$-matrix for the gravitational perturbations corresponds to the
on-shell limit of the effective action and thus, it is finite.

In the general theory, with $\La \neq 0$, we note that the parameter
$b_5$ has no effect on divergences because of the third Bianchi
identity $\na_\mu G^\mu_{\,\,\nu}=0$ and that $\na_\mu \La=0$.
Thus, there is a four-parameter $b_{1,2,3,4}$ ambiguity for
the six existing coefficients $c_{1,2,\dots 6}$. Therefore, only
two combinations of these six coefficients can be expected to be
gauge-fixing and parametrization independent. Obviously, one
of these combinations is the coefficient of $\int E_4$ defined in
(\ref{C2E4}). This
directly follows from the fact that the $\La$-term cannot affect
the four-derivative divergences, as we know from the power
counting.

Let us find the second combination of the parameters. A simple
calculation using (\ref{amb}) shows that the coefficients in the
expression (\ref{1-loop}) vary according to
\beq
&& c_1  \longrightarrow  c_1\,,
\quad \qquad
c_2  \longrightarrow  c_2+b_1\,,
\qquad \qquad
c_3  \longrightarrow  c_3 - \big(b_2+ {\textstyle \frac12}\,b_1 \big)\,,
\nonumber
\\
&&
c_4  \longrightarrow  c_4 - b_4\,,
\quad
c_5  \longrightarrow  c_5 - \big(b_1 + 4 b_2 + b_3 \big)\La\,,
\qquad
c_6  \longrightarrow  c_6 - 4 b_3\La^2\,.
\mbox{\qquad}
\mbox{\quad}
\label{trans-II}
\eeq
It is an easy exercise to show that the two gauge-fixing
and parametrization invariants which do not change under
the transformations of $c_{1,2,..,5}$ in (\ref{trans-II}), are
\beq
c_1
\qquad
\mbox{and}
\qquad
c_{inv}\,=\,c_6 - 4\La c_5 + 4\La^2 c_2 + 16\La^2 c_3\,.
\label{invs}
\eeq

The last observation is that the on-shell expressions for the
classical action and divergences have the forms
\beq
S\Big|_{on\,shell}
&=& \frac{2\La}{\ka^2} \int d^4 x \sqrt{-g}\,,
\nonumber
\\
\Ga^{(1)}_{div}\Big|_{on\,shell}
&=&
\frac{1}{\ep}\,
\int d^4x\sqrt{-g}\left\{
c_1 E_4
+ c_{inv}\right\}\,.
\label{1-loop-onshell}
\eeq
In these two functionals there are only invariant quantities. This
feature forms the basis of the so-called on-shell renormalization
group equation, to be discussed below. An additional small detail
is that $c_{inv}$ does not change if we replace $E_4$ in
(\ref{1-loop-onshell}) by the square of the Riemann tensor.

\subsection{Gauge-fixing dependence in higher-derivative models}
\label{IIQG-sec3.2}

Adding more derivatives in the classical action, the
gauge-fixing and parametrization dependence in the divergent part
of the effective action becomes smaller. Among all higher-derivative
models of QG, the unique non-trivial example is the fourth-derivative
model  (\ref{action4der}). Let us first consider this model
following \cite{a}.

In the four-derivative theory, the formula analogous to (\ref{amb})
has the form
\beq
\Ga^{(1)}_{div}(\al_i) \,-\, \Ga^{(1)}_{div}(\al^{0}_i)
\,\,=\,\,
\frac{1}{\ep}\,
\int d^4x\sqrt{-g}\,\,f_{\mu\nu}\,\,  \vp_{(4)}^{\mu\nu},
\label{amb-4der}
\eeq
where $\,f_{\mu\nu}=f_{\mu\nu} ({\alpha}_{i})\,$ is
an unknown tensor depending on ${\alpha}_{i}$ and
\beq
\vp_{(4)}^{\mu\nu}
\,=\,\frac{1}{\sqrt{-g}}\,\frac{\de S_{HD}}{\de g_{\mu\nu}}
\label{vp4}
\eeq
are the equations of motion for the fourth-derivative gravity.

To find $\,f_{\mu\nu}$, let us remember that the fourth-derivative
quantum gravity is a renormalizable theory. Therefore, all three of
the expressions $\Ga^{(1)}_{div}(\al^{0}_i)$, \
 $\Ga^{(1)}_{div}(\al_i)$ and ${\varepsilon}_{(4)}^{\mu\nu}$
have dimension 4, as the classical action. Since the divergencies
in (\ref{amb-4der}) are local functionals, $\,f_{\mu\nu}$ is a
dimensionless tensor. Then the only possible choice is
\beq
f_{\mu\nu} ({\alpha}_{i})
\,=\, g_{\mu\nu} \,f({\alpha}_{i}) ,
\label{f_mn}
\eeq
where $f({\alpha}_{i})$ is an arbitrary (can be defined only by
explicit calculations) dimensionless function of the set of
parameters of gauge fixing and metric parametrization. Thus, the
gauge/parametrization dependence of the divergent part of
effective action is controlled by the ``conformal shift'' of the
classical action
\beq
\Ga^{(1)}_{div}({\alpha}_{i})
\,-\,
\Ga^{(1)}_{div}(\al^{0}_i)
\,\,=\,\,
f({\alpha}_{i}) \int d^4 x \, \,
g_{\mu\nu}\,\frac{\delta S}{\delta g_{\mu\nu}}.
\label{conf shift}
\eeq
In the case of the conformal model, the {\it r.h.s.} of this equation
simply vanishes, i.e., in purely conformal, Weyl-squared,
gravity theory, the divergences of the effective action do not
depend on ${\alpha}_{i}$ because the classical action satisfies
the Noether identity for the conformal invariance.
For the general model (\ref{action4der}), the $C^2$, $E_4$ and
$\cx R$ terms in the action do not contribute to the {\it r.h.s.} of
(\ref{conf shift}). Then, the gauge- and parametrization dependencies
are defined by the Einstein-Hilbert, cosmological and $R^2$ terms.
It is easy to get
\beq
\Ga^{(1)}_{div}({\alpha}_{i})
\,-\,
\Ga^{(1)}_{div}(\al^{0}_i)
\,\,=\,\,
f(\alpha_{i}) \int d^4x\sqrt{- g}\,
\Big\{ \frac{2\omega}{\lambda}\cx R
\,-\,\frac{1}{\ka^2}(R+4\Lambda)\Big\}.
\mbox{\qquad}
\label{gauge dep 4der}
\eeq
The divergent
coefficient of the $\cx R$ term depends on the gauge fixing,
the same is true for the coefficients of the Einstein-Hilbert
and cosmological terms. At the same time, there are two
gauge-invariant combinations of these coefficients.

The easiest part is the gauge and parametrization dependence of
the counterterms in superrenormalizable models with more than
four derivatives, both polynomial or nonlocal.  It is easy to see
that these models do not have such dependencies. To get this
result, we note that formula (\ref{amb-4der}) is valid for all such
models, both polynomial and nonlocal, with trading of
$\vp_{(4)}^{\mu\nu}$ to the variational derivative of the
corresponding action $\vp^{\mu\nu}$. As we have seen above,
according to the power counting arguments, the divergences are
given by local expressions with four, two or zero derivatives of the
metric. On the other hand, in all superrenormalizable models, the
equations of motion $\vp^{\mu\nu}$
have more than four derivatives of the metric. Thus, the non-zero
{\it r.h.s.} of  (\ref{amb-4der}) is incompatible with the locality
of the function $f_{\mu\nu}$, proving the statement about the
universality of renormalization
\cite{highderi,SRQG-betas,CountGhosts}.

\section{One-loop divergences in quantum GR}
\label{secQG1.5}

The derivation of one-loop divergences in quantum GR has great
historical \cite{hove} and practical importance. This calculation
is a starting point for many other developments, in many different
models, including pure QG, models of more and more complicated
versions of pure QG, gravity coupled to quantum matter, etc. For
these reasons, the review on perturbative QG should include this
calculation and the list of the most important extensions and
corresponding references.

In the rest of this section, we repeat the original derivation of
divergences in pure QG from the classical paper \cite{hove}.
Since this is not complicated, we shall also include the nonzero
cosmological constant, as it was done in \cite{ChristDuff}. The
standard calculation uses the background field method based
on (\ref{bfm}), the heat-kernel expansion and the Schwinger-DeWitt
technique \cite{DeWitt65} (see also \cite{OUP} for a detailed
introduction).

The bilinear expansion of the action (\ref{EH}) is given in
Eq.~(\ref{EH-expand}), and the gauge-fixing term
with the two gauge fixing parameters $\al$ and $\be$  is given by
(\ref{qg-EH-gf}). The ghost action can be easily obtained from
(\ref{ghostM}), but we postpone this part until fixing the values
of $\al$ and $\be$. For this, we rewrite (\ref{qg-EH-gf}) as
\beq
S_{gf}
&=&
\frac{1}{\al} \int d^4x \sqrt{-g}\,\,\,
h^{\mu\nu}\Big[
g_{\mu\al} \na_\nu \na_\be
- \be \big(g_{\mu\nu} \na_\al \na_\be
- g_{\al\be} \na_\mu \na_\nu \big)
\nn
\\
&&
+\,\be^2 g_{\mu\nu} g_{\al\be}\cx \Big] h^{\al\be}.
\label{EH-gf-gen}
\eeq
Adding this expression to (\ref{EH-expand}), we require that the sum
includes the minimal operator $H_{\mu\nu,\al\be}$,
\beq
S_{EH}^{(2)} + S_{gf}
&=&
\frac{1}{2}\int d^4x \sqrt{-g}\,\,
h^{\mu\nu} H_{\mu\nu,\al\be}\, h^{\al\be}\,,
\nonumber
\\
H_{\mu\nu,\al\be}
&=&
K_{\mu\nu,\al\be}\cx + M_{\mu\nu,\al\be},
\label{bilin-EH-H}
\eeq
where $K_{\mu\nu,\al\be}$ and $M_{\mu\nu,\al\be}$ are $c$-number
operators. This is achieved for $\al=2$ and $\be=1/2$. After that, we
arrive at the expression (\ref{bilin-EH-H}) with
\beq
K_{\mu\nu,\al\be}
&=&
\frac{1}{2}\Big( \de_{\mu\nu,\al\be} - \frac12\, g_{\mu\nu}g_{\al\be}\Big),
\label{M-in-GR}
\\
M_{\mu\nu,\al\be}
&=&
R_{\mu\al\nu\be} + g_{\nu\be}R_{\mu\al}
- \frac{1}{2}\big( g_{\mu\nu}R_{\al\be} + g_{\al\be} R_{\mu\nu} \big)
- \frac{1}{2}\,R
\big( \de_{\mu\nu,\al\be} - \frac12 g_{\al\be} g_{\mu\nu} \big)\,.
\nonumber
\eeq
It is easy to see that the matrix  $2K_{\rho\si, \al\be }$
is equal to its own inverse,
\beq
\Big( \de_{\mu\nu,\al\be} - \frac12\, g_{\mu\nu}g_{\al\be}\Big)
\Big( \de^{\al\be,\rho\si} - \frac12\, g^{\al\be}g^{\rho\si}\Big)
\,\,=\,\,
\de_{\mu\nu ,}^{\quad \rho\si}\,.
\label{Kinv}
\eeq

On the other hand,
$\Tr \ln \big(2K_{\rho\si, \al\be }\big)$ does not contribute
to the divergences (e.g., in the dimensional regularization) since
this operator has no derivatives. Thus, regarding the divergences,
\beq
\Tr \ln \big(H_{\rho\si, \al\be }\big)
&=&
\Tr \ln \big(2K_{\mu\nu,}^{\quad \rho\si}\, H_{\rho\si, \al\be }\big)
\,=\,
\Tr \ln \big(H'_{\rho\si, \al\be }\big)
\nonumber
\\
&=&
\Tr \ln \big(\de_{\mu\nu,\al\be } \cx + \Pi_{\rho\si, \al\be}\big).
\label{bilin-pronto}
\eeq

A small calculation gives
\beq
{\hat \Pi}\, = \,\Pi_{\rho\si, \al\be}
&=&
2K_{\mu\nu,}^{\quad \rho\si}\, M_{\rho\si,\al\be}
\,=\,M_{\mu\nu,\al\be}.
\label{EH-Pi}
\eeq
It is evident that Eq.~(\ref{bilin-pronto}) enables one to use the
standard Schwinger-DeWitt formula for the operator
\beq
{\hat H}
\,=\,{\hat 1}{\Box}+2{\hat h^\mu}\na_\mu +{\hat \Pi}.
\label{operator1}
\eeq
For this, we need to define
\beq
{\hat S}_{\mu\nu}\,=\,\big[\na_\nu , \na_\mu\big] \hat{1}
+ (\na_\nu{\hat h}_\mu-\na_\mu{\hat h}_\nu)
+ ({\hat h}_\nu{\hat h}_\mu-{\hat h}_\mu{\hat h}_\nu)
\label{Smn}
\eeq
and
\beq
{\hat P} \,=\,{\hat \Pi}
+ \frac{\hat 1}{6}\,R - \na_\mu{\hat h}^\mu
- {\hat h}_\mu{\hat h}^\mu.
\quad
\label{P-II}
\eeq
The divergent part of the one-loop effective action is an integral
aver the ``magic'' $a_2$ coefficient,
\beq
\Ga^{(1)}_{div}
&=&
-\,\frac{\mu^{n-4}}{\vp}\,\int d^nx\sqrt{-g}\,\tr {\hat a}_2\big|,
\label{gendivs}
\eeq
where
\beq
{\hat a}_2\big|\,=\,{\hat a}_2(x,x)\,=\,
\frac{{\hat 1}}{180}\,(R_{\mu\nu\al\be}^2-R_{\al\be}^2+{\Box}R)
+ \frac12 {\hat P}^2+\frac16 ({\Box}{\hat P})
+ \frac{1}{12}{\hat S}_{\mu\nu}^2.
\mbox{\qquad}
\quad
\,\,
\label{f113} 
\eeq
In our case, we have a simple situation because ${\hat h}_\mu=0$.
Thus,
\beq
{\hat P} = {\hat \Pi} + \frac{{\hat 1}}6\,R
\qquad
\mbox{and}
\qquad
{\hat S}_{\mu\nu} = \big[\na_\nu , \na_\mu\big].
\label{PS}
\eeq
A simple calculation gives
\beq
&& {\hat P}
\,=\, P_{\mu\nu,\al\be} \,=\, {\hat K}_1 + {\hat K}_2 - \frac12 {\hat K}_3
 - \frac{5}{12}{\hat K}_4  + \frac14 {\hat K}_5,
\qquad \,\,\,\,
\mbox{where}
\nonumber
\\
&&
{\hat K}_1 = R_{\mu\al\nu\be},
\qquad\,\,\,\,\,\,
{\hat K}_2 = g_{\nu\be}R_{\mu\al},
\qquad \,\,\,\,
{\hat K}_3
= g_{\mu\nu}R_{\al\be} + g_{\al\be} R_{\mu\nu},
\nonumber
\\
&&
{\hat K}_4 =  \de_{\mu\nu,\al\be}\,R,
\qquad
{\hat K}_5 = R \,g_{\al\be}\, g_{\mu\nu},
\nonumber
\\
&&
\mbox{and}
\qquad \,\,\,\,
{\hat S}_{\la\tau}
\,=\,
\big[S_{\la\tau}\big]_{\mu\nu,\al\be}
\,=\,
- \,2 R_{\mu\al\la\tau}\,\,g_{\nu\be}.
\label{PSK}
\eeq
For the sake of compactness, in all of these expressions, we
assume automatic symmetrization over the couples of indices
$(\mu\nu)$ and $(\al\be)$.
It is easy to get the following multiplication table:
\beq
&&
\tr {\hat K}_1 \cdot {\hat K}_1 = \frac12\,R_{\mu\nu\al\be}^2
,\quad
\tr {\hat K}_2 \cdot{\hat K}_2 = \frac32\,R_{\mu\nu}^2 + \frac14\,R^2
,\quad
\tr {\hat K}_3 \cdot {\hat K}_3 = 8R_{\mu\nu}^2 + 2R^2,
\nonumber
\\
&&
\tr {\hat K}_4 \cdot {\hat K}_4 = 10R^2
,\quad
\tr {\hat K}_5 \cdot {\hat K}_5 = 16R^2
,\quad
\tr {\hat K}_1 \cdot {\hat K}_2 = - \frac12\,R_{\mu\nu}^2,
\nonumber
\\
&&
\tr {\hat K}_1 \cdot {\hat K}_3 = 2 R_{\mu\nu}^2
,\quad
\tr {\hat K}_1 \cdot {\hat K}_4 = - \frac12 R^2
,\quad
\tr {\hat K}_1 \cdot {\hat K}_5 = R^2
,\quad
\tr {\hat K}_2 \cdot {\hat K}_3 = 2 R_{\mu\nu}^2,
\nonumber
\\
&&
\tr {\hat K}_2 \cdot {\hat K}_4 = \frac52\, R^2
,\quad
\tr {\hat K}_2 \cdot {\hat K}_5 = R^2
,\quad
\tr {\hat K}_3 \cdot {\hat K}_4 = 2R^2
,\quad
\tr {\hat K}_3 \cdot {\hat K}_5 = 8 R^2,
\nonumber
\\
&&
\tr {\hat K}_4 \cdot {\hat K}_5 = 4 R^2
,\quad
\mbox{and}
\quad
{\hat S}_{\la\tau}
\cdot
{\hat S}_{\la\tau}
\,=\,
-6\,R_{\mu\nu\al\be}^2.
\label{K-tab}
\eeq
Substituting these values into Eq.~(\ref{f113}), we get
\beq
\tr \Big(
\frac{1}{2}\, {\hat P}\cdot {\hat P}
+ \frac{1}{12}\, {\hat S}_{\la\tau}\cdot {\hat S}^{\la\tau}\Big)
\,=\,
R_{\mu\nu\al\be}^2 - 3\,R_{\mu\nu\al\be}^2
+ \frac{59}{36}\,R^2
 + \frac{26}{3}\,R\La + 20 \La^2.
\qquad
\label{SP2}
\eeq
In this and subsequent expressions, we ignore the total derivative
term $\cx R$. The complete expressions can be found, e.g., in the
original publication \cite{JDG-QG}, including for the general
parametrization of quantum metric (\ref{bgf}) .

For the ghost action, we obtain
\beq
\frac{\de \chi^\mu}{\de h_{\rho\si}}
\,=\,
\de^{\mu\la , \rho\si}\na_\la \,-\,\frac12\,g^{\rho\si}\na^\mu,
\qquad
R_{\rho\si ,}^{\quad \nu}
= - \de_\rho^{\,\nu}\na_\si - \de_\si^{\,\nu}\na_\rho.
\nonumber
\eeq
Then
\beq
M_\mu^{\,\,\nu}
&=&
\frac{\de \chi^\mu}{\de h_{\rho\si}}\,R_{\rho\si ,}^{\quad \nu}
\,=\, - \,\big(\de_\mu^{\,\nu}\cx \,+\, R_\mu^{\,\nu}\big).
\mbox{\qquad}
\label{M-GR}
\eeq
This is a minimal vector operator, that enables one to use the
standard Schwinger-DeWitt formula (\ref{gendivs}). For the
commutator, we get
\beq
{\hat S}_{\la\tau}
\,=\,
\big[{\hat S}_{\la\tau}\big]_{\al , \be}
\,=\,-\,
R_{\al\be\la\tau}.
\mbox{\qquad}
\label{Sgravghost}
\eeq
Thus,
\beq
\tr \Big(
\frac12\, {\hat P}\cdot {\hat P}
+ \frac{1}{12}\, {\hat S}_{\la\tau}
\cdot {\hat S}^{\la\tau} \Big)_{ghost}
\,=\,
-\frac{1}{12}\,R_{\mu\nu\al\be}^2
+ \frac32\,R_{\mu\nu\al\be}^2
+ \frac29\,R^2.
\label{SP2ghost}
\eeq
Finally, replacing all the expressions above into (\ref{f113}) and
taking into account that, in the $h_{\mu\nu}$ sector,
$\tr {\hat 1} = \de_{\mu\nu\al\be}\de^{\mu\nu\al\be}=10$ and, in
the ghost sector, $\tr {\hat 1}_{ghost} = 4$, we arrive at the famous
result of \cite{hove},
\beq
&&
\Ga^{1,\,total}_{div} \,\,=\,\,
\frac{i}{2} \Tr\Ln {\hat H} \,-\, i\Tr\Ln {\hat H}_{ghost}
\label{GRdivs}
\\
&&
\quad
= - \,\frac{\mu^{n-4}}{\ep}
\int d^nx \sqrt{-g}\,
\Big\{
\frac{53}{45}\,R_{\mu\nu\al\be}^2 - \frac{361}{90}\,R_{\al\be}^2
+ \frac{43}{36}\, R^2 + \frac{26}{3}\,R\La + 20 \La^2
\Big\}
\nonumber
\\
&&
\quad
= - \,\frac{\mu^{n-4}}{\ep}
\int d^nx \sqrt{-g}\,
\Big\{
\frac{53}{45}\,E_4 + \frac{7}{10}\,R_{\al\be}^2
+ \frac{1}{60}\, R^2 + \frac{26}{3}\,R\La + 20 \La^2
\Big\},
\nonumber
\eeq
where, as usual, $\ep = (4\pi)^2(n-4)$,
and $\mu$ is the renormalization parameter.

As we already know from section~\ref{IIQG-sec3.1}, the
coefficient $\frac{53}{45}$ of the Gauss-Bonnet term is
invariant, while other coefficients can be modified by
changing the gauge-fixing conditions or the parametrization
of the quantum metric. The second invariant coefficient
in (\ref{invs}) can be easily found by using the on shell
conditions $R_{\mu\nu}= -\La g_{\mu\nu}$ and $R = - 4\La$.
The result of this operation is
\beq
c_{inv}\,=\, -\frac{58}{5}\,\La^2.
\label{invs-tHV}
\eeq
This means, the invariant, on shell, version of (\ref{GRdivs}) is
\beq
\Ga^{1,\,total}_{div}\bigg|
\,\,=\,\,
= \,- \,\frac{\mu^{n-4}}{\ep}
\int d^nx \sqrt{-g}\,
\Big\{\frac{53}{45}\,E_4 \,-\, \frac{58}{5}\,\La^2 \Big\},
\label{GRdivs-onshell}
\eeq
\vskip 1mm

To conclude this section, let us make a few observations.
\vskip 1mm

i) \ The reader could note that in the formulas in the previous section,
e.g., in (\ref{gauge dep 4der}), we used $\int d^4x\sqrt{-g}$, while
in this section, the more complicated integration rule
$\int d^nx \mu^{n-4}\sqrt{-g}$ was used. The point is that the two
formulas are equivalent when it concerns the divergences. The
expression with $\int d^4x$ is just shorter and the one with
$\int d^nx$ may be more useful, e.g., for deriving the beta function
in the Minimal Subtraction Scheme of renormalization.
\vskip 1mm

ii) \ One may be curious why we identify the square of the
Riemann tensor in the formula for invariant part of divergences
in Eq.~(\ref{1-loop-onshell}) with the Gauss-Bonnet topological
invariant $E_4$ and not with the square of the Weyl tensor. This
is an important question and we shall give an extensive answer to it.

This issue requires addressing the nonlocal finite contributions
corresponding to the logarithmic UV divergences. One can calculate
these nonlocal terms directly (see, e.g., \cite{avramidi-86,bavi90}
for the heat-kernel approach that can be used for this purpose),
or using Feynman diagrams, including in the theory of massive
fields \cite{apco} (see also \cite{Codello,Omar-FF4D,OUP}).
In the massless limit, the nonlocal logarithmic terms correspond
the higher-derivative divergences (\ref{1-loop}) and have the form
\cite{OUP}
\beq
\Ga^{(1)}_{fin,\,HD}
&=&
\int d^4x\sqrt{-g}\,\bigg\{
c_1\,R_{\mu\nu\al\be}\ln \Big( -\,\frac{\cx}{\mu^2} \Big)
R^{\mu\nu\al\be}
\nn
\\
&&
\quad
\,+\,\, c_2 R_{\al\be}\ln \Big( -\,\frac{\cx}{\mu^2} \Big)R^{\al\be}
\,+\, c_3R \ln\Big( -\,\frac{\cx}{\mu^2} \Big)R\bigg\}.
\label{1-loop-nonloca}
\eeq
Let us note that one can also derive the nonlocal form factor for the
Einstein-Hilbert term \cite{Codello,Omar-FF4D}, regardless there is
an ambiguity even in the nonlocal representation of the proper
action of GR \cite{Barvinsky2003}. However, what we need is just
the expression (\ref{1-loop-nonloca}), which is the physical
reflection of the divergences.

The correspondence between divergences and nonlocal terms holds
independent on the choice of gauge fixing and parametrization, such
that the transformations  (\ref{trans-II}) apply to the coefficients
of the logarithmic terms in  (\ref{1-loop-nonloca}). As we know from
section \ref{secQG1.3}, all three parts of this expression affect the
propagation of the spin-2 or spin-0 modes of the metric. The
ambiguity  (\ref{trans-II}) affects this propagation and, in
particular, can eliminate all of it. As we know from the discussion
of propagator in the general models (\ref{act Phi}) with an extra
term (\ref{GBterm}), the unique part of the expression which does
not affect the  propagation of the spin-2 or spin-0 modes of the
metric, is the generalized Gauss-Bonnet term
\beq
\Ga^{(1)}_{GB,\,nonloc}
\,&=&\,
-\,\,c_1\,\int d^4x\sqrt{-g}\,\bigg\{
R_{\mu\nu\al\be}\ln \Big( -\,\frac{\cx}{\mu^2} \Big)
R^{\mu\nu\al\be}
\nn
\\
&&
-\,\,\, 4 R_{\al\be}\ln \Big( -\,\frac{\cx}{\mu^2} \Big)R^{\al\be}
\, +\, R\ln\Big( -\,\frac{\cx}{\mu^2} \Big)R\bigg\}.
\label{GB-log}
\eeq
This means, the invariant coefficient $c_1$ in (\ref{invs}) should
be attributed to the $E_4$ and not to the Weyl-squared term. In
case of QG based on GR, this coefficient is $53/45$. There may be
further contributions from matter fields, which do not depend on
the gauge fixing in the QG sector of the theory.
\vskip 1mm

iii) \ As one can see already from the power counting arguments,
the renormalization of the the Einstein-Hilbert term in QG based
on GR is possible only owing to the nonzero cosmological
constant $\La$. This feature holds beyond the one-loop order and
can be seen as a general result. However, this
statement corresponds to the most universal, logarithmic,
divergences. The quadratic divergences of the Einstein-Hilbert
type are possible even for $\La=0$. However, we know that these
divergences depend on the choice of regularization and, therefore,
the physical results which one can obtain from quadratic
divergences always have uncertain physical sense. This issue was
discussed in more detail, e.g., in \cite{OUP}.

\section{On shell renormalization group in quantum GR}
\label{secQG1.5.3}

The formulation of renormalization group running in QG is a
complicated problem because  renormalizable and superrenormalizable
models of QG always have massive ghosts and other massive degrees
of freedom, such that the physical sense of
the the running (and quantum effects, in general) in these models is
not clear. The reason is that, loop contributions is these models
come from the functional integral over massless mode of the
gravitational field and, also, over the massive modes. It is a
standard assumption that the contributions of massive fields vanish
at the energy scale below their masses. In the fourth derivative QG
this expectation was formulated, for the first time, in \cite{frts82}.
Nowadays, it is partially supported by the calculations in
semiclassical theories \cite{apco,Omar-FF4D} and in the toy model
of QG \cite{324}.
On the other hand, in QG all massive modes
have masses of the Planck order of magnitude (see the discussion
of this issue in \cite{ABSh2}). Thus, the physical applicability of
the renormalization group running in fourth- and higher-order
models of QG is restricted by the UV domain, with energies
\textit{above} the Planck scale.

Thus, despite the running of parameters of the actions in all
renormalizable QG models might have great importance from the
theoretical perspective, its application to any kind of  physics is
unclear. Since the universal QG theory in the IR is quantum GR
\cite{don}, we can try to explore the running using this model.
However, the use of renormalization group to explore the running
in the quantum GR is a non-trivial issue because the theory is not
renormalizable. Assuming that all massive degrees of freedom
decouple  in the IR (below the Planck scale) and the physically
interesting running is in the region below the Planck scale of
energies, we arrive at the subject of effective approach to QG,
which is treated in a separate Section of this Handbook.

Let us consider a version of renormalization group running in QG
which enables one to avoid the mentioned difficulties. This version
of renormalization group is not perfectly well defined, but it gives
an idea of what we can expect from the running.
The on shell version of renormalization group uses the expressions
(\ref{1-loop-onshell}) for the classical action and one-loop
divergences on the classical equations of motion. The on shell
divergences are universal, i.e., do not depend on the parametrization
of quantum metric and on the gauge-fixing choice. Ignoring
the topological term $c_1E_4$, the aforementioned expressions can be
used to perform the renormalization \textit{on shell} and,
consequently, to achieve the renormalization group running.

Let us define the dimensionless parameter
\beq
\ga = \ka^2 \La.
\label{80}
\eeq
In terms of this parameter, the classical action and the
one-loop counterterm (both on-shell) have the form
\beq
S_{EH} \Big|_{on-shell}
&=&
- \, \frac{1}{\ka^4} \int d^n x \sqrt{-g} \,\mu^{n-4}\,(- 2 \ga),
\nn
\\
\De S^{(1)} \Big|_{on-shell}
&=&
\frac{1}{\vp}\,\cdot\,\frac{1}{\ka^4} \int d^n x \sqrt{-g}
\,\mu^{n-4}\,
\Big( - \frac{58}{5}\ga^2 \Big).
\label{79}
\eeq
As these two expressions have an identical dependence of the
metric, we can remove the divergence by making the
renormalization transformation,
\beq
\ga_0 &=& \mu^{n-4}\Big(\ga - \frac{29}{5\vp}\,\ga^2 \Big) \,,
\label{reno}
\eeq
from what immediately follows the general $\be$-function
in $n$ spacetime dimensions,
\beq
\label{88-n}
\mu\,\frac{d\ga}{d\mu}
\,=\, -\,(n-4)\,\ga
\,-\,\frac{29}{5(4\pi)^2} \,\ga^2.
\eeq
The standard  $\be$-function for $\ga$ can be obtained in the
limit $n\to 4$ and we arrive at the renormalization group equation
that corresponds to an asymptotically free theory,
\beq
\label{AFga}
\mu\,\frac{d\ga}{d\mu}
\,=\, \be_{\ga}
\,=\, \,-\,a^2\,\ga^2,
\qquad
a^2=\frac{29}{5(4\pi)^2}\,.
\eeq
The solution of this equation can be easily found if we set the
initial value at some scale, $\ga(\mu_0)=\ga_0$,
\beq
\label{AFcc}
\ga(\mu)
\,\,=\,\,\frac{\ga_0}{1 + a^2 \ga_0 \ln (\mu/\mu_0)}\,.
\eeq
The physical interpretation of this solution meets several
difficulties, so let us present some points discussing this subject.

1. \ It is possible to identify $\mu$ with one or another
physical parameter, in different physical situations. As far as
$\ga$ is the dimensionless cosmological constant, the standard
application of the solution should be in cosmology and then
the natural choice of the scale is the Hubble parameter
\cite{Babic,CC-nova,Babic-setting}.

2. \ The cosmological constant that enters the definition (\ref{80})
is not the one which is responsible for the observed accelerated
expansion of the Universe. The observable density of the cosmological
constant $\rho_\La^{obs}=\La/(8\pi G)$ is a sum of the vacuum quantity  $\rho_\La^{vac}=\rho_\La^{vac}$ and the induced density
$\rho_\La^{ind}$ coming from, e.g., the electroweak phase transition.
Both summands are many orders of magnitude greater than the observed
quantity $\rho_\La^{obs}$. For this reason, the numerical value of
$\ga_0$ may be small, but it is not that small as some reader might
think. This issue was discussed in detail in \cite{UEA-RG}, with the
consideration based on the unique effective action formalism
\cite{Vilk-Uni,DeWitt-ea,TV90}. This interesting approach is discussed
in a  separate Chapter of this Section and does not fit the present review.
However, we shall briefly discuss the differences between the two
universal equations for $\ga(\mu)$ below.

3. \ Another interesting point is that, if  $\ga_0$ is sufficiently
small
are negligible and this equation can be regarded as non-perturbative.
To see this, it is sufficient to pick up the power counting in
the quantum GR, as it was presented above.

To end this section, let us compare the running (\ref{AFcc}) with
the one in the effective approach to quantum GR with cosmological
constant \cite{UEA-RG}\footnote{I am grateful to Breno
Giacchini and Tib\'{e}rio de Paula Netto for the stimulating
discussions of this issue.}  (see also earlier work \cite{TV90} with
the same result but without effective interpretation). In the
effective approach, the higher derivative terms should be neglected,
but without these terms, the invariants (\ref{invs}) cannot be
constructed. The way out is to use the Vilkovisky-DeWitt scheme of
unique effective action, that does not depend of parametrization or
gauge fixing by construction \cite{Vilk-Uni,DeWitt-ea}. The
calculations in this case give \cite{TV90}
\beq
&&
\Ga^{1,\,total}_{div} \,\,=\,\,
- \,\frac{\mu^{n-4}}{\ep}
\int d^nx \sqrt{-g}\,
\Big\{
\frac{53}{45}\,E_4 + \frac{121}{60}\,R_{\al\be}^2
- \frac{29}{60}\, R^2 + 8\,R\La + 12 \La^2\Big\}.
\nn
\\
\label{GRdivs-UEA}
\eeq
This is different from (\ref{GRdivs}), but it is easy to check that
the on shell expression (\ref{GRdivs-onshell}) is the same. But in
this case, we do not need to rely on the on shell version.
Neglecting all fourth derivative terms in (\ref{GRdivs-UEA}) and
using the standard formalism, one gets the renormalization group
equations for the constants in the action (\ref{EH}),
\beq
&&
\mu\frac{\text{d}}{\text{d}\mu}\frac{1}{\ka^2}
\,=\,
\frac{8\La}{(4\pi)^2}\,,
\label{RG-ka}
\\
&&
\mu\frac{\text{d} \La}{\text{d}\mu}
\,=\,
- \frac{2 \La^2\ka^2}{(4\pi)^2}\,.
\label{RG-La}
\eeq
In this case, the equation for the dimensionless combination
(\ref{80}) has the same general form (\ref{AFga}), but this time
with the coefficient
\beq
\label{AFga-effect}
a^2=\frac{10}{(4\pi)^2}\,.
\eeq
This change of the beta function illustrates the role of the
physical interpretation in QG. The difference between the two
cases is that in (\ref{AFga}) we do not ignore the terms with
four derivatives and get the invariant beta function by using the
on shell condition. On the contrary, in (\ref{AFga-effect}) we
follow the effective approach, ignore higher derivative terms
and use the Vilkovisky's unique effective action to provide
an invariant result.

Both approaches look satisfactory and both are not perfect.
Which one is better? In a ``normal'' physical model the answer
would be probably given by experiment, but in QG this is not
an option.

\section{One-loop divergences in other models of QG}
\label{secQG1.xxx}

There is a vast number of publications on the one-loop calculations
and we do not pretend to make a complete review or even list the most
relevant of these works (including the ones of the present author)
here. In what follows, I will separate those papers which were first
of the kind for the most relevant models.

The first of these works concerned the one loop divergences in
quantum GR coupled to quantum matter. Already in the first paper
with one-loop calculations, 't Hooft and Veltman derived also the
divergences for QG coupled to the minimal scalar field \cite{hove}.
It is interesting that the case of quantum GR coupled  to the
nonminimal quantum scalar $\phi$ (nonminimal means the presence
of the term $\xi R\phi^2$) can be reduced to the calculations with
a minimal scalar by the change of variables, i.e., the conformal
transformation \cite{BKK}. However, the correspondence with the
direct calculation \cite{spec} required much greater effort and was
achieved only two decades later in \cite{KamStei}. This example shows
importance and complications related to the choice of parametrization.
The calculations in gravity-vector (Abelian and nonabelian) and
gravity-fermion models were done almost at the same time as
\cite{hove} by Deser, van Nieuwenhuisen et al in \cite{dene}. Let
us note that more general combinations of
quantum fields including QG were explored in supergravity, but
since there is a special Section of this Handbook about supergravity,
we do not need to discuss this part here. The important general
result of the mentioned calculations is that, for matter-gravity
systems, the one-loop divergences do not vanish on shell
\cite{hove,dene}.

The one-loop divergences were calculated in other models of QG,
including fourth-derivative QG with the action (\ref{action4der}) and
the polynomial superrenormalizable model (\ref{gaction}). The case
of the theory (\ref{action4der}) is treated in full detail in another
Chapter of this Section of our Handbook and we will not discuss
the technical details here. From the general perspective, this
calculation was relevant for several reasons. There is a growing
understanding that the construction of QG which is not restricted
to the low-energy domain, cannot be successful without higher
derivative and the same is true in the semiclassical theory. Then,
one of the ``traditional'' expectations is to advance in the problem
of ghosts and instabilities by exploring quantum corrections and,
at the first place, the logarithmic contributions
\cite{Tomboulis-77,salstr}. As we already know, these  contributions
are directly related to the UV divergences

Historically, the calculation of divergences in the fourth-derivative
QG started in the work \cite{julton}, using Feynman diagrams
and separating massless and massive internal lines. This way to
make calculations was never used after that (up to our knowledge)
as all subsequent calculations were performed by technically more
simple heat-kernel (or Schwinger-DeWitt) technique, adapted for
the four-derivative operators in \cite{frts82} and
later on, in a more systematic way, in \cite{bavi85}. However, the
diagrammatic approach of the pioneering work of Julve and Tonin
\cite{julton} may be, by the end of the day, rather important. The
reason is that, using diagrams enable one to separate the graphs
with only massless internal lines from the ones which have the
lines of the massive degrees of freedom (i.e., ghosts, in the present
case). This method can be used to explore
the IR limit of the quantum corrections in higher derivative theories.
As we already mentioned above, in this area we have only the general
statement about the universality of quantum GR as an effective theory
of QG in the IR \cite{frts82,don} and the calculations of decoupling
in the models of massive matter fields \cite{apco,324} (see further
references therein) that can be used as toy models for the problem
of decoupling massive degrees of freedom in QG. Let is stress that
completing the program of decoupling in the higher derivative models
would be important for improved understanding of QG as a whole.

The next important step was done in the seminal paper \cite{frts82}
by Fradkin and Tseytlin. This was the first publication were the
relevance of the weight operator (\ref{weight-4der}) for quantum
contributions was noted, there was also the first heat-kernel
treatment of the fourth derivative operator and of the non-minimal
vector operator, the first discussion of the IR decoupling in higher
derivative theories, and the first observation concerning the
difference between general and conformal QG models. We can say,
that \cite{frts82} paved the way for the development of  a large part
of QG in the subsequent decades. Furthermore, the correct result
for the general and conformal versions of the fourth-derivative QG
models was obtained in \cite{avbar86} and \cite{AMM}, respectively.
Finally, there was a confirmation of the correctness of these results
in \cite{Weyl,Gauss}, where another interesting aspect of QG was
addressed. Much earlier, in the paper \cite{capkim}, it was noted
that the Gauss-Bonnet term $\int E_4$ may play a nontrivial role in
QG. The point is that the topological nature of this term becomes
badly seen in the perturbative approach, especially if one uses the
dimensional regularization. In the previous sections we saw that the
topological term does not contribute to the propagator, in any
dimension $n$. However, for $n \neq 4$ it does affect the vertices.
The question posed in  \cite{capkim} was whether it is true that the
contributions of these vertices cancel in the divergent part of
effective action, or even in the finite part of it. The answer is that
the divergences really do not depend on the $\int E_4$ term, and
this dependence takes place only in the local finite terms.

Finally,  let us mention  the derivation of the one-loop divergences
in the polynomial, superrenormalizable models of QG
\cite{SRQG-betas}. From the technical side, these calculations
require the generalized Schwinger-DeWitt technique developed
by Barvinsky and Vilkovisky \cite{bavi85}.

\section{Concluding discussion}
\label{secQG1.7}

Quantum gravity is intended as a theory to be most relevant in the
vicinity of the singularities, i.e., at extremely high energies. The
idea to formulate QG in a way of perturbative QFT, that proved
efficient in other theories, is the first that comes to mind when one
decides how to quantize gravity. In this sense, the perturbative
approach is the most basic part of the whole QG program.

To a great extent, the current situation in perturbative QG is
similar to the one in QG, in general. We have many approaches
to incorporating the quantum effects of gravity within the
well-established perturbative formalism of quantum field theory.
On the other hand, there is no perfect model of quantum
gravity. The theory based on GR is a promising candidate
to describe low energy effects (see the corresponding Section
on effective QG), but it has serious problems with
non-renormalizability in the UV. From another side, fourth
derivative model is renormalizable and enables one to make
controllable calculations at any energy scale. However, the
spectrum of particles in this theory includes nonphysical massive
ghosts. The instabilities of classical solutions, generated by these
ghosts, look non-avoidable.  However, our experience in
cosmology shows that these instabilities show up only in the
regions with the typical energies are of the Planck order of
magnitude. The role of the ghosts and the emergence of
instabilities in the superrenormalizable QG models were not
explored at the same level of quality, but it is expected that
the output may be, qualitatively, the same.

We can draw two main conclusions about the main problem
of perturbative QG. First, this problem is the conflict between
renormalizability and the lack of physical unitarity (which does
not reduce to the unitarity of the $S$-matrix \cite{Asorey-2018})
or instabilities of classical solutions owing to the presence of
ghosts. Second,
the resolution of this conflict does not look impossible, but it
will (most likely, at least) require new ideas and new approaches.
According to DeWitt and Molina-Paris  \cite{BDW_CMP},
long ago W. Pauli remarked ``It will take somebody really smart''
to construct a quantum theory of gravity. After many decades
that passed after this prediction, we are in a position
to say that it will take somebody really smart to find modifications
of gravity explaining how Nature prevents the ghosts to emerge.
The opinion of the present author is that the problem of ghosts
shows up in the perturbative approach, but its solution will, most
likely, be found only beyond this framework.

From the historical perspective, one of the main impacts of QG
to quantum field theory was the development of the classical and
quantum theory of gauge fields, that helped to develop the Yang-Mills
theory and gave a theoretical basis to the Standard Model of Particle
Physics. In this respect, it is sufficient to mention the work of
De Witt \cite{DeWitt-67}.  It is quite possible that the further
progress in QG will be based on the flux of ideas in the opposite
direction, that is using methods which already exist in field theory,
but are not sufficiently familiar to the QG community.

As one can learn from the variety of approaches presented in this
Handbook, we have many theories of QG. The main problem is perhaps
not the shortage of the theories, but that none of these theories can
be currently tested in experiments and/or observations. However, the
increasing volume and improving quality of the observational data in
astrophysics and cosmology may help to close this gap, someday.
Thus, at least part of the nowadays theoretical developments
in QG may find their use in the future.

\section*{Acknowledgements}

The author is partially supported by Conselho Nacional de
Desenvolvimento Cient\'{i}fico e Tecnol\'{o}gico - CNPq (Brazil),
the grant 303635/2018-5 and by Funda\c{c}\~{a}o de Amparo \`a
Pesquisa de Minas Gerais - FAPEMIG, the project PPM-00604-18.


\end{document}